# Ferroelectric Materials for Synaptic Transistors and Their Neuromorphic Applications

Zexin Wang[1]

[1]School of Materials Science and Engineering, Shanghai University, Shanghai,200444, China

**ABSTRACT**

After more than a hundred years of development, ferroelectric materials have demonstrated their strong potential to people, and more and more ferroelectric materials are being used in the research of ferroelectric transistors (FeFETs). As a new generation of neuromorphic devices, ferroelectric materials have attracted people's attention due to their powerful functions and many characteristics. This article summarizes the development of ferroelectric material systems in recent years and discusses the simulation of artificial synapses. The mainstream ferroelectric materials are divided into traditional perovskite structure, fluorite structure, organic polymer, and new 2D van der Waals ferroelectricity. The principles, research progress, and optimization for brain like computers of each material system are introduced, and the latest application progress is summarized. Finally, the scope of application of different material systems is discussed, with the aim of helping people screen out different material systems based on different needs.

**1. Introduction**

In the age of intelligence, the amount of information that needs to be stored and processed is exploding. The traditional computer based on von Neumann architecture cannot avoid data transmission between the processing unit and the storage unit when processing these data, which will cause the problem of computing speed decrease and energy consumption increase[1]. Therefore, it is proposed to build new computing networks by learning information processing patterns in the human brain to address the limitations of existing computer architectures. The human brain has $10^{11}$ neurons but requires only 20W of power[2], Neurons in the brain connect to each other through synapses, which are the key unit of information processing in the human brain[3]. At first, it was hoped that existing complementary metal oxide semiconductor (CMOS) circuits could be used to run artificial neural networks[4], but simulating a single synaptic function would require several transistors[5], which would greatly reduce energy efficiency and increase the difficulty of integration. Therefore, the research of high performance and low power devices with synaptic structure and function becomes the key to realize efficient neural networks[6].

Over the past few years, researchers have made breakthroughs in using various memristors to create artificial synapses and neurons[7–10]. Typically, memristors have a simple two-terminal structure characterized by an internal resistance state that changes with applied voltage and current and preserves this "state", the so-called memory function[11]. Similarly, the human brain behaves similarly: synapses change their strength of connection (synaptic weights) in response to surrounding stimuli.

Therefore, memristors can be used in synaptic devices and are very suitable for integrating high density neural networks due to their simple structure. However, in the integration of cross grid array, the calculation accuracy of array will be affected by serious current interference, which leads to inevitable error[12]. In addition, which limits their ability to transmit signals and update weights. Fortunately, three-terminal synaptic devices are effective at avoiding these problems[13]. The source drain of the three-terminal synapse as the input terminal is independent, its read and write processes are completely separated, and can simultaneously input signal processing and weight change. Ferroelectric transistor (FeFET) is a kind of three-terminal synaptic device, which has the advantages of lossless readout, low power consumption and high working speed, and is considered as an excellent solution of three-terminal synaptic device. For example, wang et al. demonstrated a multifunctional FeFET that can integrate logic, memory computing, photoelectric logic, and non-volatile computing into a single transistor. The gate of this transistor can also be used as a signal input, showing the powerful potential of FeFET in synaptic devices[14].

The most central material in FeFETs is the ferroelectric material. Ferroelectric materials have developed into an important class of functional materials. Ferroelectricity was first discovered in Rochelle salts[15], however, the unstable and water-soluble nature has greatly limited their applications and research. During World War II, the discovery of the ferroelectricity of the chalcogenide material barium titanate (BTO) gave a great impetus to the development of ferroelectric materials for extreme applications, followed by the emergence of many different materials and structures[16]. The characteristic of ferroelectric materials is that spontaneous polarization can be generated within a certain temperature range and the polarization properties are regulated by an external electric field. This property can be observed through the polarization electric field curve（Figure 1h）. In addition, ferroelectrics usually have basic properties such as high dielectric constant, piezoelectricity, and pyroelectricity. Today, many types of materials based on ferroelectric transistors have been developed, including: Perovskite structure（Figure 1b）（Figure 1c）, fluorite structures（Figure 1a）, organic polymers（Figure 1d）, and two-dimensional ferroelectric materials（Figure 1e-g）.

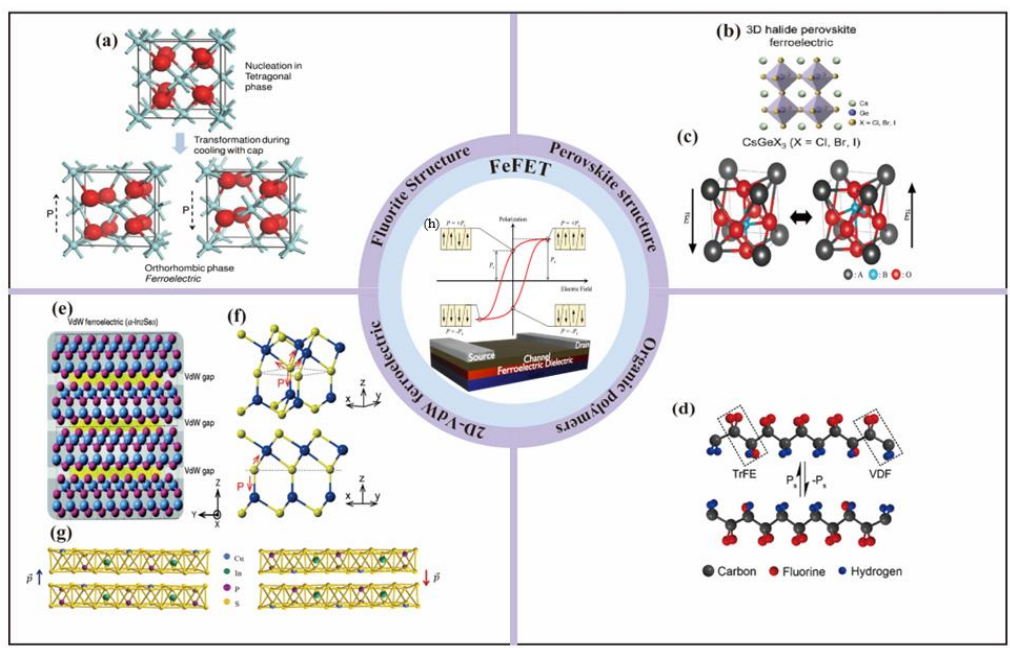

Figure 1(a). Fluorite structure ferroelectric undergoes a transformation from tetragonal phase to rhombohedral phase, with the bottom indicating the change in polarization states. (b). Novel 3D halide perovskite ferroelectric: CsGeX3 (X = Cl, Br, I). (c). Transformation of the traditional BTO ferroelectric structure under an electric field. (d). Structure of P(VDF-TrFE) with opposing dipole moments. (e). Schematic diagram of the typical van der Waals (VdW) ferroelectric crystal structure. (f). Crystal structure of ferroelectric α-In2Se3 with different dipole directions, where yellow and blue spheres represent Se and In atoms, respectively. (g). Side view of the CIPS crystal structure with two opposing polarization directions. The polarization direction is indicated by arrows.

## 2.Synaptic properties and ferroelectric field-effect transistors

So far, various neuromorphic devices with different functions have been realized. These functions are almost all realized by imitating the information transmission method of the brain[17,18]. For example, transistors simulate biological synapses through electrical pulse signals[19]. Therefore, it is necessary to understand the information transmission mode and transmission structure in the brain. When the brain processes information, it relies on synapses connecting the presynaptic and postsynaptic neurons[20]. To achieve advanced neuromorphic devices, simulating the function of synapses becomes essential. The synapse structure comprises the presynaptic membrane, synaptic cleft, and postsynaptic membrane[21]. This intricate system facilitates the transmission and processing of synaptic information, where each neuron connects to others through thousands of synapses[22]. The connection strength of pre-neurons and post-neurons is called synaptic weight, which can be adjusted by synaptic plasticity, and synaptic weight is an index to measure the functional characteristics of synapses in neural networks. The change of synaptic weight is called synaptic plasticity, and this plastic function that realizes the adjustable connection strength between neurons is the basis of biological learning, reasoning and other behaviors. Synaptic plasticity can be divided into short-term synaptic plasticity and long-term synaptic plasticity according to the length of duration. Short-term synapse generally lasts from a few milliseconds to a few seconds. During this period, synapses will experience short-term potentiation (STP) and short-term depression (STD). Short-term synaptic plasticity plays an important role in neural signal transmission and information processing, such as the realization of paired-pulse potentiation (PPF) and paired-pulse depression (PPD). Long-term synaptic plasticity includes long-term potentiation (LTP) and long-term depression (LTD). Generally, LTP dominates the learning of new information, while LTD removes unnecessary old information and maintains a balanced relationship between learning and forgetting. There are also spike-time dependent plasticity (STDP) and spike-rate dependent synaptic plasticity (SRDP). The SRDP rule means that the synaptic weight will change with the frequency of the peak, while STDP shows that the synaptic weight is determined by the time difference before and after the peak. STDP, one of the Hebbian learning rules, well explains the relationship between LTP and LTD: when the pre-synaptic spikes arrives before the post-synaptic spikes (Pre-Post Spiking), it can cause LTP; when the post-synaptic spikes arrives before the pre-synaptic spikes (Post-Pre Spiking), Can cause LTD.

## 3 Development of ferroelectric materials and ferroelectric transistors

Ferroelectric materials boast a rich historical background, with Valasek pioneering the measurement of ferroelectricity in Rochelle salts as early as 1921[23]. Following this landmark study, the presence of ferroelectricity has been subsequently identified in numerous perovskite oxides[24–29]. In 1957, Bell Laboratories proposed utilizing the distinctive properties of ferroelectric materials to modulate the surface conductivity of semiconductors[30]. Previously, the existence of ferroelectricity has been verified in BTO and Pb(Zr,Ti)O$_3$ (PZT). These two materials are still the most representative perovskite ferroelectric materials today, and play a role in many fields, such as ferroelectric capacitors, ferroelectric memories, sensors and detectors, etc[16]. Advancements in thin-film growth techniques have streamlined the production of FeFETs, allowing the integration of numerous ferroelectrics into transistors. Following perovskite ferroelectric materials, organic ferroelectric materials entered the scene. In 1978, Davis et al. validated the ferroelectric properties of polyvinylidene fluoride (PVDF)[31]. In the 1980s, a PVDF derivative, polyvinylidene fluoride-trifluoroethylene [P(VDF-TrFE)], demonstrated ferroelectric properties[32]. Meanwhile, in 1993, Ishiwara introduced a novel adaptive learning neuron circuit. This circuit utilized the conductance change in the FFET channel to mimic the synaptic weight alterations in the human brain, marking the integration of FeFET into neuromorphic device studies[33].

In 2007, Intel introduced HfO2, a crucial material, into semiconductor manufacturing for its excellent CMOS compatibility as a high-k dielectric. The discovery of ferroelectricity in HfO2 in 2011 led to its rapid application in FeFET technology[34]. In recent years, the remarkable properties of 2D materials have garnered significant interest[35–38]. Moreover, theoretical predictions indicate the potential for ferroelectricity in certain 2D semiconductors characterized by asymmetric crystal structures[39]. Two-dimensional ferroelectric materials that have been discovered include: CuInP$_2$S$_6$ (CIPS)[40], MoTe$_2$[41,42], MoS$_2$[43,44] and α-In$_2$Se$_3$[39,45]. Two-dimensional ferroelectrics exhibit the polarization characteristics typical of general ferroelectrics. What adds to their appeal is the excellent optical and electrical properties inherent in two-dimensional materials. Their combination with traditional ferroelectric materials enhances their functionality, unlocking more powerful capabilities[45,46].

### 3.1 FeFET based on perovskite material

Perovskite ferroelectric materials encompass perovskite halides and perovskite oxides. The development of perovskite halide ferroelectric materials is still in its early stages, and their physical properties remain unclear[47]. Despite this, studying perovskite halide ferroelectricity is deemed necessary. Firstly, as an emerging ferroelectric material, its low density and soft structure align with the requirements for flexible and miniaturized devices. Furthermore, the ferroelectricity observed in perovskite halides may be linked to their superior optoelectronic properties, holding significant potential for optoelectronic synaptic devices. Presently, perovskite halide materials are in the initial exploration phase concerning neuromorphic computing and non-volatile storage. Research on two-terminal neural devices with a simpler structure is progressing more rapidly than the exploration of three-terminal devices[48]. For example, Yao et al. prepared the perovskite ferroelectric material BCPB by unit-transmutation method (figure 2a), which has unique ferroelectric photovoltaic characteristics and has both ferroelectric and semiconductor properties. BCPB exhibits a stable spontaneous polarization of 6.5 μCcm-2 (figure 2c), with a clear zero-bias photocurrent density (figure 2b) and high on/off switching ratio of current, which surpasses the well-known ferroelectric semiconductor BiFeO$_3$, which allows it to read out information nondestructively as an photovoltaic non-volatile memories. At the same time, the unique layered structure makes the device have strong stability when storing information, and still maintains a strong polarization ability after $10^8$ bipolar switching cycles(figure 2e)[49].

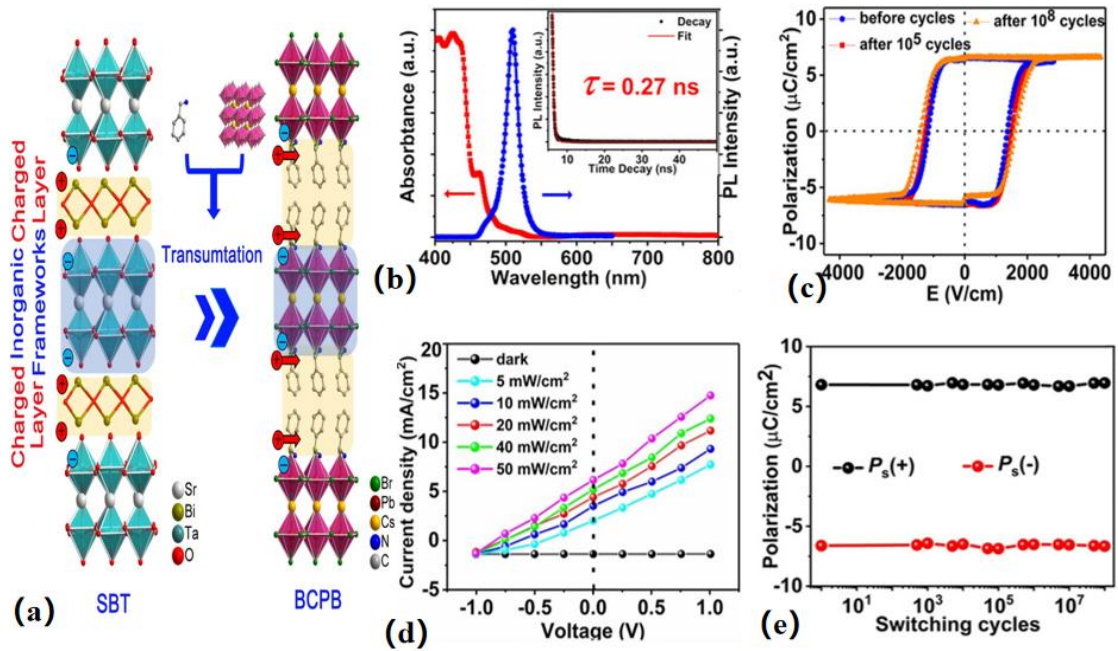

Figure 2(a) Transformation of fatigue-free ferroelectric SBT into the layered hybrid perovskite BCPB. (b) Optical absorption and photoluminescence spectra of BCPB, with an inset showing the time-resolved photoluminescence spectrum of BCPB. (c) Hysteresis loops for BCPB before and after 108 cycles. (d) Photocurrent density of BCPB under 405 nm illuminations at different power levels. (e) Variation of Ps (polarization) versus the number of switching cycles.

Perovskite oxide materials were the initial choice for the fabrication of FeFETs, leveraging their notable advantages of good stability and large remanent polarization. Dating back to the early 20th century, Yoon et al. pioneered the use of perovskite oxides in constructing tunable synaptic arrays of FeFETs. These synaptic arrays, employing SBTO materials, demonstrated the capability for weighting and computation in artificial neural networks. Subsequent research efforts successfully fine-tuned the synaptic performance of these devices[50,51].

To meet the demands of emerging neuromorphic device applications, perovskite oxide ferroelectrics are frequently integrated with high-performance optoelectronic materials. Moreover, in the heterogeneous structure of a single perovskite oxide, lattice mismatch can impact interface quality, and 2D materials exhibit a synergistic effect with the oxide interface. The integration of 2D layered materials and perovskite oxide ferroelectrics demonstrates anticipated functionality and enhances device performance[52]. For example, Du et al[53]. combined BTO ferroelectric film and single-layer $MoS_2$ to prepare phototransistors that can be used in neuromorphic visual sensors(figure 3b). The phototransistor can not only regulate polarization by conventional electricity, but also can reverse by light-induced ferroelectric plan. Taking advantage of the response of $MoS_2$ to different wavelengths of light, the multi-level optical storage properties are realized, and the preprocessing of neuromorphic vision within the sensor is realized. The research team demonstrated the image acquisition and detection functions of the device on a 5×5 array. The array's operation involved the application of both optical pulses and electrical pulses (5 V for 100 ms) to simulate writing and erasing procedures. The conductance readings were taken 10 s after the light pulse stimuli. Initially, before any optical stimuli, all devices in the array exhibited a low-conductance state (figure 3a I). Subsequently, an "X" image was inscribed onto the vision sensor array using an optical pulse train with specific parameters (wavelength of 450 nm, number of 30, intensity of 10 mW/cm$^2$, width of 100 ms, and interval of 100 ms) (figure 3a II). The study explored two distinct operational modes: "electrical erasing - optical writing" (figure 3a III) and "optical erasing - electrical writing" (d). Furthermore, the image of "Y" was written by employing optical pulses with specific parameters (wavelength of 532 nm, number of 30, intensity of 10 mW/cm2, width of 100 ms, and interval of 100 ms), and retention characteristics were investigated after 5 min (figure 3a IV). Another instance of writing the image "Y" involved optical pulses with different parameters (wavelength of 650 nm, number of 30, intensity of 10 mW/cm2, width of 100 ms, and interval of 100 ms), and the retention characteristic was measured after 300 s (figure 3a V). The transistor has a high switching ratio and good volatility, and the image recognition rate is obviously improved after preprocessing. Puebla et al. demonstrated a single-layer MoS2-based BTO-FeFET with a mobility comparable to that of a high-performance hexagonal boron nitride device, demonstrating an example of a combination of 2D materials and complex oxides[54]. As illustrated in Figure 3cI, Liu et al. presented a dual-gate ferroelectric MoS2 Field-Effect Transistor (FeFET) within a complete van der Waals structure. The study utilized the ferroelectric gate dielectric NBIT and few-layer graphene as barrier-free contacts[55]. Clockwise hysteresis was observed with either a top ferroelectric gate or SiO2 back gate, attributed to interface charge dynamics and gate coupling. Dual-gate modulation transformed the clockwise hysteresis into counterclockwise ferroelectric hysteresis under specific conditions（Figure 3cII，Figure 3cIII）. The research also assessed the device's data retention and endurance characteristics for memory operations（Figure 3cIV-VI. In a separate study, Luo et al. employed layered transition metal sulfide WS2 as a conductive channel and PZT film as a ferroelectric gate medium. As depicted in Figure 3d, the memory transistors, thus prepared, could regulate conductance by voltage and light, simulate short- and long-term synaptic plasticity, and implement brain-like learning rules driven by light information[56].

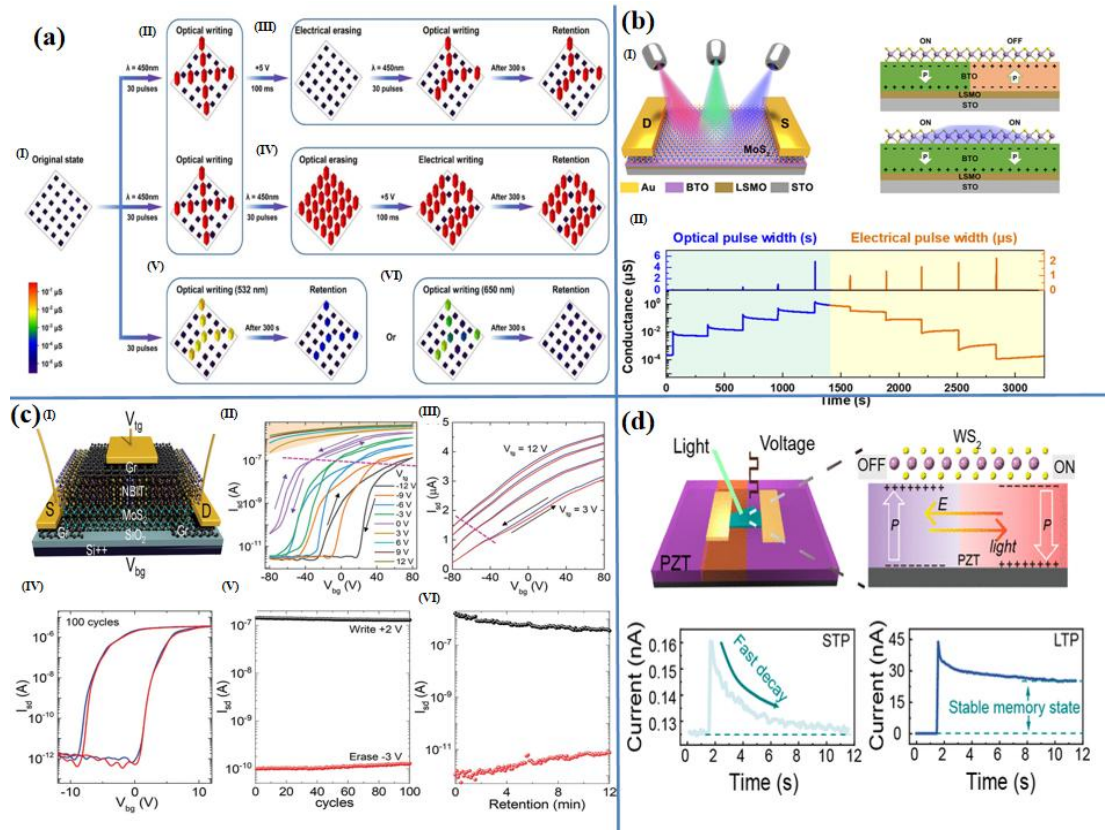

Figure 3 (a). Schematic diagram of the detection and writing principles of the visual sensing array. (b). Structural diagram, with the planning process heavily dependent on the intensity of light exposure. Multiple memory characteristics are achieved using the wavelength properties of light. (c). Ⅰ Transfer characteristic curve under $V_{bg}$ modulation at $V_{tg}$. Ⅱ Hysteresis behavior changing with $V_{tg}$ gate bias. As $V_{tg}$ increases, the transition point continuously moves toward more negative $V_{bg}$, as indicated by the pink dashed line. Ⅲ Transfer curves at higher $V_{tg}$ are linearly plotted to better illustrate hysteresis behavior. Ⅳ Transfer curve after 100 cycles. Ⅴ Durability test of 100 write and erase operations. Ⅵ Retention characteristics of the device.(d). The upper part shows a schematic diagram of the device structure and its principles of optoelectronic conductance modulation. The lower part illustrates light-triggered STP (Short-Term Potentiation) and LTP (Long-Term Potentiation) behaviors.

### 3.2 FeFET based on Fluorite-structured ferroelectrics

Ⅰ In 2011, the discovery of ferroelectricity in fluorite-structured metal oxides ($HfO_2$, $ZrO_2$) marked a significant breakthrough[57–59]. This revelation, holding great promise for semiconductor devices, garnered widespread attention upon publication. Unlike previously reported oxide materials such as BTO, which faced challenges in practical production due to incompatibility with CMOS technology, HfO2 emerged as a notable candidate. Notably, HfO2 has been employed as a high-k dielectric in CMOS technology since 2007[60], elevating expectations for its commercial applications. A year after the identification of HfO2 ferroelectricity, a ferroelectric transistor with a channel length of 28nm was reported in 2011[61]. Subsequently, in 2017, Dünkel et al. demonstrated a 22nm ferroelectric fully depleted silicon-on-insulator (FDSOI) device[62], underscoring the advantages of fluorite-structured ferroelectric materials in terms of integration and CMOS compatibility.

In classical ferroelectric material systems, ferroelectrics often exhibit an asymmetric crystal structure, as exemplified by BTO. The fluorite structure is no exception, where oxygen ions within the unit cell are offset to generate an asymmetric structure. However, the polarization mechanism resulting from this partial offset is not entirely identical[63]. $HfO_2$, with its fluorite structure, showcases a variety of surprising domain and domain wall structures[64]. Despite advancements, the ferroelectric mechanism of HfO2 remains incompletely understood, suggesting unexplored potentials6[65]. Furthermore, HfO-based materials offer the advantage of ultra-thinness. Studies have reported that the ferroelectricity of HZO films deposited directly on silicon substrates can be maintained down to a thickness of 1 nm[66,67].

**Origin of Thin Film Ferroelectricity**
$HfO_2$ and $ZrO_2$ are polymorphic, demonstrating distinct crystal structures in different environmental conditions. Taking HfO2 as an example, at normal temperature and pressure, it exists in a stable monoclinic phase (m phase). As the temperature increases, this phase transforms to a tetragonal phase (t phase, at 1973 K) and further transitions to a cubic phase (c phase, at 2777 K). With increased pressure, the stable crystal phase undergoes a transformation to orthorhombic i phase (OI phase) and orthorhombic ii phase (OII phase). Both $ZrO_2$ and $HfO_2$ share similar phase transition tendencies, with ferroelectric orthorhombic phases (OIII phase and OIV phase) appearing only under extreme conditions. The discovery of ferroelectricity in HfO2 and $ZrO_2$ appears to be a fortuitous occurrence, given the specific environmental conditions required for the manifestation of this phenomenon.

Fluorite structure materials, when prepared, generally do not exhibit ferroelectricity externally, and no non-centrosymmetric phase is evident in the temperature-pressure phase diagram[68,69], Consequently, HfO thin films grown through conventional methods lack ferroelectricity, with the ferroelectric phase emerging only under extreme conditions, such as high-stress manufacturing technology, suggesting a potential for ferroelectricity in the film[70]. Most current research utilizes Atomic Layer Deposition (ALD) technology to prepare hafnium-based ferroelectric films[71]. The origin of ferroelectricity in fluorite-structured oxides remains relatively complicated, lacking a perfect explanation. Schroeder et al., in conjunction with recent reports on the root cause of ferroelectricity, conducted a comprehensive analysis of fluorite-structured oxides' ferroelectricity. The analysis of ferroelectricity in $HfO_2$-based materials involved both thermodynamic and kinetic approaches, paving the way for future research[34]. However, HfO$_2$ materials face challenges in characterization, necessitating effective methods for studying the ferroelectricity of nanoscale hafnium-based thin films. In-situ observation techniques during the polarization transition process prove invaluable in this regard. For example, Nukala et al. employed direct oxygen imaging atomic microscopy to observe the interface of La$_{0.67}$Sr$_{0.33}$MnO$_3$/HZO capacitors under in-situ electrical bias, revealing the reason for the nanoscale ferroelectricity: reversible oxygen vacancy migration[63]. As shown in Figure 4a, Zhong et al. employed laser pulse deposition technology to synthesize an HZO/LSMO heterostructure on a single crystal SrTiO3 (STO) substrate. The HZO thin film could be peeled off in a solution environment, offering significant advantages for studying crystal phases on a microscopic scale（Figure 4b）. Simultaneously, it was observed that the ferroelectric phase is most stable under conditions without a substrate. This observation indicates that the stress from sinking to the bottom is unrelated to maintaining ferroelectric polarity. Consequently, independently prepared HZO thin films are beneficial for researching flexible electronic devices[72].

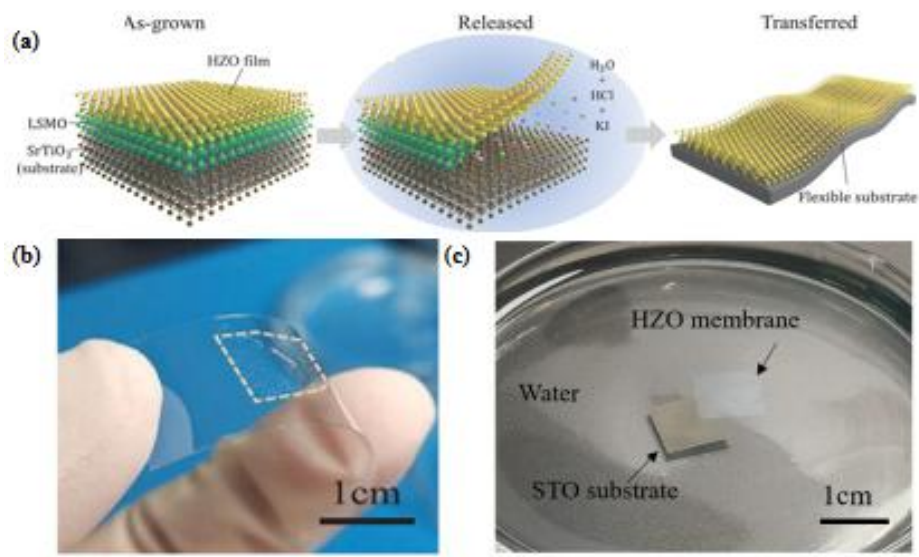

Figure 4 (a). Schematic diagram of the preparation process, where the underlying LSMO can be dissolved in the etchant to obtain independent HZO films. These films can then be transferred to a new substrate to form heterostructures and interfaces. (b). Photograph of the transfer onto a PET substrate. (c). Separation of the HZO film from the substrate in water.

The reliability and durability of stored data depend significantly on the ferroelectric properties, as data recording involves the inversion of ferroelectric domains. This necessitates the ferroelectric functional layer to maintain stable ferroelectric properties. Similar to HfO$_2$, ZrO$_2$ requires doping or annealing to induce the formation of the ferroelectric OIII phase, leading to the generation of ferroelectricity. Various factors, including dopant type and concentration, surface energy, and mechanical stress, are crucial in determining the ferroelectricity of ferroelectric thin films.

The dopants of hafnium oxide ferroelectric thin films can be divided into metals and non-metals. The reported metal elements for doping are Si, Al, La, Gd, Y, Sr, Zr, Ce, Ca, Mg, Ba, Fe, Lu, Pr[57,70,73–81]. In metal doping, the choice of metals with small or large atomic radii influences the metal-oxygen bond length, altering the crystal structure's asymmetry. Notably, small atomic radius elements like Si and Al can induce antiferroelectric properties. The untapped potential of antiferroelectric materials in energy storage and low-power devices is demonstrated by Cao and collaborators, showcasing the ultra-low power consumption (37fj/spike) and exceptional performance of antiferroelectric field-effect transistors (MNIST recognition accuracy rate 96.8%)[82].

Non-metallic doping elements, such as N, C, H[83–85], are typically introduced during the preparation or annealing processes. For instance, incomplete precursor decomposition during low-temperature ALD can lead to the unintentional introduction of these impurities. Coincidentally, in some cases, these impurities may have a positive impact on ferroelectricity [83].

As previously mentioned, the O phase is a critical factor for materials to exhibit ferroelectricity, and exploring ways to facilitate the generation of the O phase is beneficial. In hafnium oxide-based materials, both the bulk and surface energies of the m-phase and t-phase are smaller than those of the O-phase[86,87]. The size-driven phase transition from the nanostructured m-phase to the t/o-phase can make the O phase easier to generate. Therefore, constructing nanolayered structures can limit the grain size[88]. Moreover, limiting grain growth can be achieved through thinner films and lower annealing temperatures. The surface effect significantly influences the stability of HfO2-based films and the ferroelectric O phase. Ferroelectricity may diminish with increased thickness and larger grain size[89,90].

Throughout the film growth process, mechanical stress significantly restrains the volume of the grain, impacting film performance. This stress originates primarily from two sources: the top electrode and the bottom substrate. A suitably matched electrode positively influences the ferroelectric properties of the film, with TiN being a commonly used top electrode for $HfO_2$ films[91]. The substrate plays a decisive role in growth direction, where the crystal orientation, coefficient of thermal expansion (CTE), and lattice parameters of the substrate collectively influence the film's properties[92].

**Application of fluorite structure ferroelectric FeFET**

Image recognition stands as a prominent application of ferroelectric synapses, with common validations involving simple images like handwritten digit recognition. Various neural networks are employed in image recognition, each tailored to specific contexts based on their inherent characteristics. Kim et al. showcased the utilization of nano-scale $HfO_2$ films for FeFET preparation, achieving data retention and multi-level storage capabilities in transistors. These transistors exhibit strong linearity, high Gmax/Gmin, and can effectively emulate the facilitation and inhibition behaviors of synapses. Additionally, the researchers simulated an artificial neural network using the National Institute of Standards and Technology (MNIST) database, comprising 400 input neurons, 100 hidden neurons, and 10 output neurons, corresponding to 20×20 data. After 125 training iterations, the system achieved an impressive 91.1% correct rate in recognizing digits 0 to 9. This research has pioneered the implementation of neuromorphic hardware functions through the use of ferroelectric transistors[93];Kim et al. successfully employed an Al-doped $HfO_2$ film in realizing artificial synapses（Figure 5a）. The FeFET synapse device exhibits precise control over channel conductance, enabling plasticity features such as PPF and STDP. The high-durability linear curve demonstrates the synaptic device's resilience over numerous cycles, maintaining optimal performance even after 100 cycles and 14,000 pulses. Additionally, the device exhibits a significant advantage in terms of switching speed[94].

The Convolutional Neural Network (CNN) excels in recognizing complex images compared to general neural networks. However, the convolution process entails repetitive calculations, posing a substantial challenge for computing systems. Traditional computers struggle to efficiently execute such tasks, necessitating a parallel structure for integrated storage and calculation[95,96]. In response to this challenge, Kim et al. devised an integrated FeFET synapse array based on HZO ferroelectric material, enabling both non-volatile storage and access functions. This parallel computing capability aligns well with CNN requirements, significantly enhancing image recognition abilities. As shown in Figure 5b, the array effectively extracts features from input images with dimensions of 64×64 pixels, and its application in simulating CIFAR-10 demonstrates the potential of HZO-based FeFETs as neuromorphic hardware for CNNs[97].

$Ca^{2+}$ dynamics in biological synapses closely resemble the persistent photoconductivity (PPC) observed in photonic synaptic devices. Lee's research group engineered a relaxation-tunable photosynaptic device utilizing hafnium zirconium oxide and oxide semiconductors. Leveraging PPC properties akin to biological behavior, the device simulates short-term plasticity, paired-pulse facilitation (PPF), and long-term plasticity observed in synapses（Figure 5c）. Notably, under light influence, the conductance of the Indium Gallium Zinc Oxide (IGZO) channel undergoes time-dependent changes. Simultaneously, the polarization deflection of the ferroelectric layer impacts the conductance, causing the recombination of photogenerated electrons on the interface due to carrier depletion and accumulation, thereby inducing synaptic relaxation[98]. Jeon et al. introduced a method for regulating the Hafnium Zirconium Oxide (HZO) interface layer. They employed self-assembled monolayer materials to adjust the passivation layer at the HZO and $MoS_2$ interface, mitigating interface traps. They elucidated the hysteresis of the passivation technology through the modulation mechanism of electronic behavior. Furthermore, given $MoS_2$'s response behavior in the visible light range resembling the human visual system, simulations were conducted under dark and visible light conditions (Figure 5d). This resulted in the update of salient weights and multi-level conductance characteristics, achieving high switching ratios, linear weight changes, and 38 levels of conductance states[99].

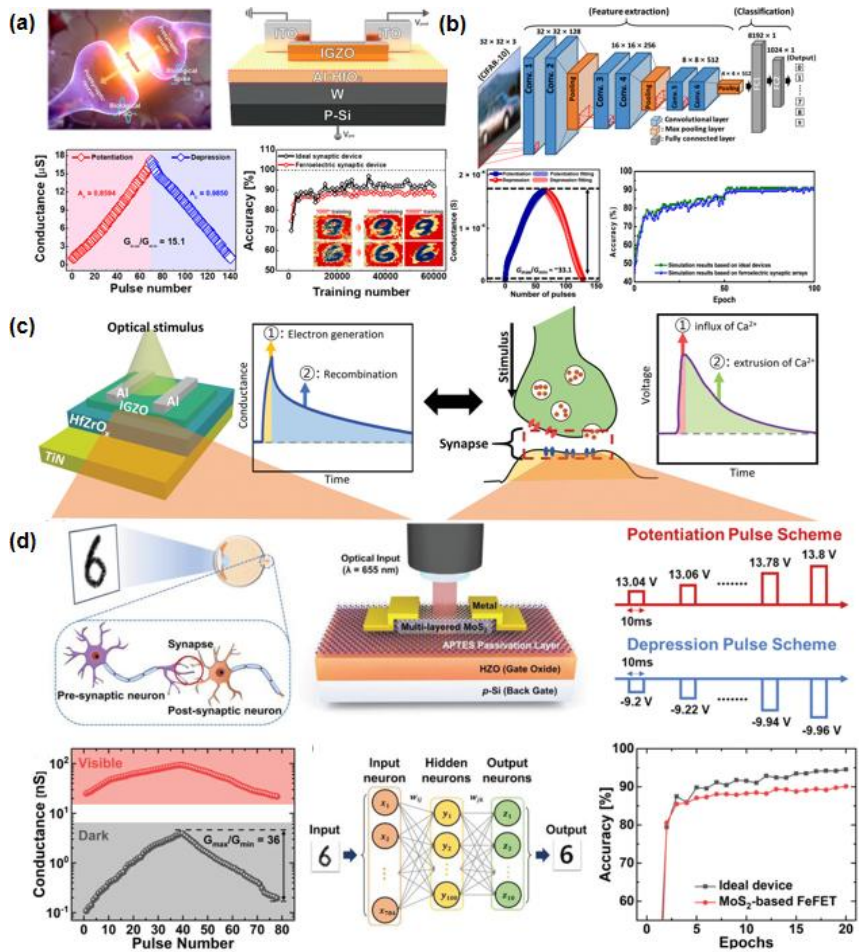

Figure5 (a). Schematic diagram of the structure of a biological synaptic device, with its linear enhancement/suppression characteristics and recognition accuracy shown below. (b). Schematic diagram of the VGG-8 network for the CIFAR-10 dataset. Below are the weight update characteristics of FeFET and a comparison of simulation accuracy with an ideal device. (c). IGZO-based light-responsive FeFET and its light-response mechanism—ionization and recombination processes of oxygen vacancies. (d). Schematic diagram of a photonic ferroelectric multi-level neural synaptic device with an APTES passivation layer. The highlighted characteristics under different conditions are shown, and the lower part of the diagram displays a comparison between the pattern recognition accuracy based on ANN simulation mode and an ideal device.

  The integration of artificial intelligence into medical applications is not a recent development. However, to cater to the needs of sudden disease detection, long-term treatment, and intelligent prosthetics, it becomes crucial to amalgamate electronic devices with flexible technology. Hafnium-based ferroelectric materials have faced limitations in flexibility due to the requirement of high-temperature annealing for ferroelectricity generation, hindering device integration. To address these challenges, researchers have explored various approaches, focusing on annealing temperature and device structure. Joh et al. implemented a low-temperature process involving microwave annealing to achieve robust ferroelectricity in Hafnium Zirconium Oxide (HZO). The results indicate that the HZO film on a flexible mica substrate maintains excellent ferroelectricity even after 10,000 cycles of bending, demonstrating good electrical properties, including a 100 ns switching speed and reliability (> $10^7$). Leveraging these experimental outcomes, a flexible device was constructed, successfully simulating synaptic behavior.[100]. Xia et al. have introduced an innovative approach to achieve flexible FeFETs by implementing a low-temperature growth process below 90 °C for ferrodielectric layers. They successfully fabricated carbon nanotube-based protrusions on flexible substrates with a 2 μm thickness, creating a tactile FeFET. This transistor exhibits excellent stability, maintaining functionality in the environment for 240 days. It boasts a dynamic range of 2000X and 360 distinguishable conductance states. To assess its performance, the array architecture was tested using the NeroSim+3.0 framework, achieving a remarkable recognition accuracy of 94.5%.[101].

### 3.3 FeFET based on organic polymer materials

While inorganic ferroelectric oxides have witnessed significant advancements, ferroelectric polymers offer distinct advantages, particularly in terms of environmental friendliness, biocompatibility, and mechanical durability. Ferroelectric polymers can be dissolved in organic solvents, enabling various liquid-phase preparation methods, including direct printing of thin films on flexible substrates. This facilitates the cost-effective and straightforward creation of large-scale synaptic networks, a feat challenging to achieve with inorganic ferroelectric thin films. Numerous polymer materials, including nylon, polyurea, liquid crystal, and certain supramolecular materials like trialkylbenzene 1,3,5-tricarboxamide, exhibit ferroelectric properties. These attributes enhance the feasibility of industrial applications in the field.[102–105].

Polyvinylidene fluoride (PVDF) and its derivatives hold a prominent position in the realm of neuromorphic devices. Renowned for their superior thermal stability and ferroelectric properties compared to typical polymers, PVDF features a basic repeating unit derived from a vinylidene fluoride monomer with a chemical structure based on CH2-CF2. In this structure, two fluorine atoms and two hydrogen atoms are chemically bonded to the main carbon chain, imparting the material with robust strength and toughness. The piezoelectric and ferroelectric characteristics of PVDF were first reported by Kawai in 1969 and 1978[31,106], respectively. In the 1980s ,the derivative trifluoroethylene was discovered. The ferroelectric polarization of this type of material comes from the strongly electronegative fluorine atoms in the VDF molecule. Under the condition of an external electric field, the main chain of the polymer rotates to cause ferroelectric inversion. Furukawa et al. explained the inversion of polymer ferroelectric materials in more detail[32]. The complete inversion process consists of three steps. Initially, there is the rapid rotation of a single molecular chain, followed by an expansion to a small range of rotation. Finally, interactions between different regions lead to the complete flipping of the entire film. The first step, involving the rotation of a single molecule, occurs within 50 ps, minimizing its impact on the ferroelectric switching speed. However, the subsequent process, analogous to the "nucleation" process observed in oxide ferroelectricity, is the primary factor limiting the speed of ferroelectric flipping.

The molecular chain of PVDF exhibits a distribution of positively charged hydrogen atoms and negatively charged fluorine atoms, resulting in a high vacuum dipole moment. The different orientations of the dipole moment lead to its polymorphic crystal structure with various molecular configurations, including the all-trans "Z" configuration (TTTT), the trans-twist-trans-twist (TGTG'), and configurations in between (TTTG). PVDF is recognized to have five phases: α, β, γ, δ, and ε phases[107], each representing distinct molecular arrangements and configurations within the PVDF material. Under the condition of applying a high external electric field, there will be a transition between the phases[108]. Among the five phases of PVDF, α-PVDF is the most stable, but it lacks ferroelectricity. The β, γ, and δ phases exhibit permanent dipole moments, with the β phase having the largest dipole moment[109,110]. The β phase is crucial for piezoelectricity and ferroelectricity in PVDF. However, during crystallization, PVDF tends to form the thermally stable α phase[111]. Many current studies focus on improving the content of the β phase in PVDF by optimizing the preparation process. One effective method is to incorporate TrFE (trifluoroethylene) to enhance the ferroelectric performance. Introducing trifluoroethylene to replace some hydrogen atoms forms polyvinylidene fluoride-trifluoroethylene polymers (P(VDF)-TrFE). The optimal ferroelectric performance is achieved when the TrFE content is in the range of 20% to 50%[112–115]. Additionally, other fluorinated monomers such as chlorofluoroethylene (CFE) and chlorotrifluoroethylene (CTFE) have been explored for improving PVDF's ferroelectric properties. Another approach to obtain stable β-PVDF involves applying tension to α-PVDF or subjecting it to electrical or optical treatments to induce the transformation of crystals into the β phase[116,117].

The production of high-quality ferroelectric polymer films is crucial for enhancing the reliability and durability of transistors. Several factors are known to influence the ferroelectricity of these films, including film thickness, dopant type and concentration, annealing conditions, and the composition of the gas environment during fabrication[118,119]. Reducing the thickness of the film to below 200 nm can significantly decrease the content of the β phase and weaken the phenomenon of spontaneous polarization[115]. Additionally, thinning the film may lead to other issues, particularly related to charge leakage. Therefore, a combination of a ferroelectric copolymer with a high-insulation dielectric and a top coercive electric field is proposed as a gate material. Xu and Xiang et al. A sandwich structure was proposed, using two AlOx interfacial layers to sandwich an ultra-thin polymer ferroelectric film, which significantly suppressed the gate leakage current and also reduced the operating voltage[112]. Furthermore, adjusting the annealing temperature of the film is crucial for maintaining the permanent dipole moment within the desired thickness range of 1 nm. Environmental factors, including temperature, humidity, atmosphere, and cooling rate, can also impact the polarization switching and domain walls of the film, consequently influencing the reliability of the device.

Various methods are employed for the preparation of polymer films, including molecular beam epitaxy (MBE), Langmuir-Blodgett (LB) film technology, and spin coating.

Molecular Beam Epitaxy (MBE): While MBE involves a relatively complicated process, it offers the advantage of low costs. Conducted under ultra-high vacuum conditions, MBE allows for precise control of molecular orientation, resulting in a regular two-dimensional stacked structure.

Langmuir-Blodgett (LB) Film Technology: LB film technology arranges immiscible molecules at the gas-liquid interface to form a molecular film, which is then transferred to a solid substrate. It can be categorized into vertical deposition and horizontal deposition methods. Additionally, a rolling method has been developed to create LB films, offering benefits such as the ability to produce ultra-thin films, control thickness, maintain temperature conditions, ensure repeatability, and achieve uniform textures with minimal defects.

Spin Coating Method: The most common method for preparing PVDF films is spin coating. This technique is characterized by a simple preparation process, low cost, and easy control. The spin coating process involves four key steps: preparing a uniform solution, dripping the solution onto the substrate, spinning the substrate to spread the solution evenly, and drying the film. The desired film thickness can be achieved by adjusting the solution concentration.

Each of these methods has its advantages and is chosen based on factors such as complexity, cost, precision, and specific requirements of the application.

**Organic polymer FeFET application**

Polymer ferroelectric materials serve as excellent coupling platforms for electrical, mechanical, and thermal energy, and their ability to be processed into thin, flexible films or fibers makes them widely applicable in miniaturized human-computer interaction devices and wearable technologies[120]. Hao et al. provided a comprehensive exploration of the unique physical properties of polymeric ferroelectric materials, offering a detailed discussion of the structure and physical mechanisms of Ferroelectric Field-Effect Transistors (FEFETs). This work contributes to a better understanding of the potential applications and behavior of polymer ferroelectric materials[121]. To reduce the power consumption of synaptic devices, Xie et al. demonstrated artificial synaptic FeFETs with P(VDF-TrFE)-wrapped InGaAs nanowires (NWs)(Figure 6c). One-dimensional materials such as NWs closely resemble the topology of tubular axons, and they also have good carrier mobility and scaling properties. The wrapping structure of NWs proved instrumental in reducing leakage and working voltage. One-dimensional NWs also demonstrated distinctive negative photoconductivity properties. Pavlov's associative learning experiments were conducted using positive and negative photoconductivity. Visible light and infrared light were leveraged to achieve linear long-term depression (LTD) characteristics. This innovative approach significantly reduced energy consumption while maintaining the accuracy of the learning process. This work showcases the potential for developing energy-efficient neuromorphic devices with advanced functionalities.[122].

Dang et al. engineered a three-terminal synaptic FeFET featuring P-type high-mobility black phosphorus coupled with a ferroelectric copolymer, P(VDF-TrFE), as the ferroelectric layer, achieving a remarkable mobility of 900 cm²V⁻¹s⁻¹ with an on/off ratio of $10^3$. Impressively, the energy consumption for a single pulse event is remarkably low, approximately 40fJ, closely resembling the efficiency of human brain synapses (10fJ). Through the modulation of ferroelectric gates and the adjustment of synaptic weights, the study simulated various synaptic behaviors, including Long-Term Potentiation (LTP), Long-Term Depression (LTD), Paired-Pulse Facilitation (PPF), and memory consolidation processes. Employing the hardware neural network for pattern recognition resulted in an accuracy rate of 93.6%[123]. Yan et al. introduce a novel ferroelectric synaptic transistor network designed to facilitate associative learning and one-step recall of entire datasets from partial information. The interplay between the external field and the internal depolarizing field governs ferroelectric creep of domain walls, delivering Hebbian synaptic plasticity with a minimal energy cost per ferroelectric synapse, measured at full and sub-femtojoule levels. This encompasses short-term memory (STM) to long-term memory (LTM), alongside pronounced peak time-dependent plasticity (STDP) and peak rate-dependent plasticity (SRDP). By utilizing a third terminal to control ferroelectric domain dynamics, the system achieves adaptive coupling between neurons, synchronously updating synaptic weights. Experimental validations, including Pavlov's dog experiments and multiassociative memory tasks, attest to the efficacy of this ferroelectric synaptic transistor network. Moreover, it demonstrates potential applications in constructing multilayer neural networks, offering a novel avenue for associative memory information processing[124].

Polymer ferroelectric materials find promising applications in intelligent sensing, with a particular focus on mimicking the functionality of human sensory organs. The skin, as a vital sensory organ, facilitates the perception and transmission of information from the surroundings. Human sensory organs are intricately designed, featuring stimulus-sensitive cells such as visual photoreceptors, olfactory and chemoreceptors responsible for taste, and mechanoreceptors contributing to hearing. Notably, these cells exhibit synapse-like connections to afferent neurons, forming a sophisticated network for sensory signal processing. Harnessing the properties of polymer ferroelectric materials opens avenues for developing artificial sensory systems that emulate the capabilities of human sensing organs. This has implications for advancing intelligent sensing technologies in various applications. The brain's synapses enable parallel information processing with high energy efficiency, mirroring the intelligent functions performed by neurons in sensory organs. In sensory organs, neurons with synaptic structural connections engage in intelligent processes such as adaptation, filtering, amplification, and memory before transmitting information to the brain. Consequently, electronic devices need to possess the capability to sense, store, and process information, aligning with the concept of sensor-storage-computing integrated devices. Additionally, as an organ with flexibility, the skin imposes a requirement for flexible devices. Polymer FeFET synaptic devices emerge as a solution that can simultaneously meet these requirements. Merkel cells, which are mechanosensitive cells in the skin forming synapses with afferent neurons, create structures known as Merkel cell-neurite complexes (MCNCs). Lee et al. present a flexible artificial synapse receptor designed to mimic the synaptic connections found in MCNCs. The receptor utilizes organic ferroelectric field-effect transistors that leverage tribocapacitive coupling effects for gate switch control. The tribocapacitive coupling effect is initiated when tactile stimuli are applied to the device substrate's surface, as illustrated in Figure 7a. In this process, the polarization deflection of the ferroelectric layer induces changes in conductance, thereby modulating the post-synaptic current signal and enabling the self-powered transmission of tactile information. The device's synaptic functionality supports multifunctional preprocessing of the output signal. Tunability of synaptic weights in prominent tactile senses is achieved by adjusting the composition of a nanocomposite ferroelectric layer comprising barium titanate nanoparticles and polyvinyldifluoride-trifluoroethylene (P(VDF-TrFE)). Additionally, the sensory memory of a 2 × 2 sensor array was simulated to recognize the number and sequence of touches without requiring additional signal processing after stimuli cessation[125]. Lee, Jang, et al.[126] proposed a smart tactile electronic skin with learning ability based on a FeFET array, capable of sensing and learning multiple tactile information simultaneously. Upon applying haptic pressure to the original top gate, the ferroelectric material undergoes polarization deflection, causing the channel conductance to change based on the size and number of pressure peaks at the top(Figure 7b). In a continuous input test of 10,000 pulses, the device exhibited high cycle stability and achieved long-term inhibition and enhancement of synaptic function. The device proved to be extremely robust. A 4×4 array was constructed for analog learning and handwritten digit recognition, achieving an accuracy of 99.66% even under 10% noise conditions.

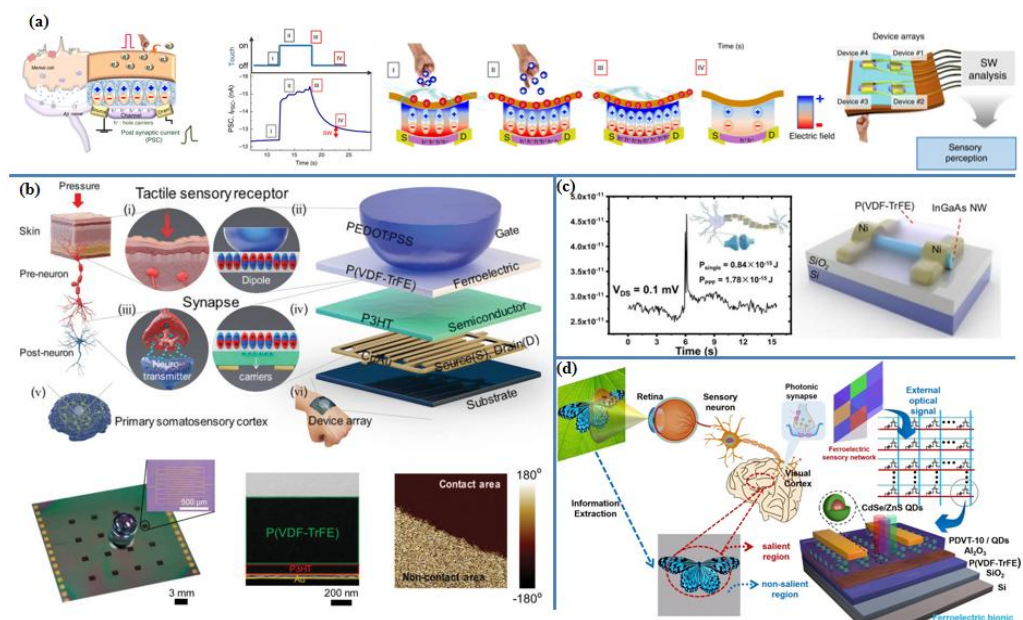

Figure 6 (a). Schematic diagram of the device corresponding to Merkel cells and MCNC

structures, along with their working principles. (b). Structure of the ATFES device. Biological tactile perception system diagram: (i) tactile sensor, (iii) synapse, (v) primary somatosensory cortex of the brain. Corresponding artificial tactile learning electronic skin based on Fe-FET devices: (ii) artificial tactile reception system with a dome-shaped elastic gate electrode and a ferroelectric layer underneath, (iv) artificial synapse with excess carriers corresponding to neurotransmitters generated by ferroelectric dipoles, (vi) an array of artificial tactile learning devices on the skin, equivalent to cortical neural networks. The bottom images include a 4 × 4 array of ATFES devices, cross-sectional TEM images, and PFM images of the P(VDF-TrFE) layer at the boundary. The inset is an optical microscope image of the ATFES device.(c)Ferroelectric P(VDF-TrFE)-capped InGaAs nanowire artificial synapses exhibit ultra-low power consumption and synergistic photoelectric modulation

(d). Schematic diagram of the biological visual perception system and an individual artificial ferroelectric biomimetic vision hardware.

In addition to electronic skin based on tactile perception, the application of organic ferroelectrics also focuses on neuromorphic systems, such as visual perception and multifunctional perception. Optical signals possess stronger information transmission capabilities than electrons. Light of different wavelengths significantly increases the information capacity it can carry, and its propagation speed is faster, leading to enhanced information transmission rates during calculations. In the future, light is expected to offer more advantages in communication and computing, displaying strong development potential. Selective recording of visual information can significantly enhance the efficiency of image information recognition. Yu et al[127]. proposed an artificial visual perception array based on ferroelectric synapses, capable of simulating biological visual information processing in a single device while selectively recording and suppressing UV information at high current decibels (figure 7d). The research team introduce a ferroelectric-based programmable bionic vision hardware that emulates biological visual information processing within a single device. The ferroelectric layer imparts tunable, nonvolatile, and programmable synaptic states to the device, allowing selective recording of external visual information. Consequently, the artificial visual perception arrays created with these devices demonstrate viability in selectively attending to UV optical information based on different polarization directions. Leveraging the linear drain current and high wavelength dependence, the device facilitates the implementation of a neural network for image recognition and classification. The accuracy significantly improves from 69.7% to 95.7% with selective attention. In addition, many researchers have conducted research based on photoelectric synaptic transistors. Li et al. proposed a wearable artificial synaptic device with optical and electrical modulation modes. The synthesized device has an ultra-fast operating speed of 30 ns and an Ultra-low power consumption. Artificial synapse realizes associative learning under composite optoelectronic modulation. The ultra-low power artificial synapse has good plasticity under different bending strains[128]. Jiang et al. designed and fabricated an asymmetric FeFET with a MoS$_2$ channel, and obtained robust electrical and optical cycling with a large switching ratio of $10^6$ and 9 different resistance states. At the same time, the operation of memory mimics the behavior of synapses in response to light pulses of different intensities and numbers[129].

In the application of multifunctional perception, researchers have also made a lot of innovative work. The supersensory neuromorphic device that can recognize, remember and learn stimuli that are difficult for humans to perceive has caused considerable interest in the research of interactive intelligent electronics.

Kim et al. introduce an AI magnetoreceptive synapse inspired by birds' magnetocognitive abilities employed for navigation and orientation. The proposed synaptic platform relies on an array of ferroelectric field-effect transistors featuring air-levitated magnetically interacting top gates. The suspended gate, composed of an elastomer composite with a conductive polymer-laminated layer of superparamagnetic particles, undergoes mechanical deformation under a magnetic field. This deformation allows control over the magnetic field-dependent contact area between the suspended gate and the underlying ferroelectric layer. The remanent polarization of the ferroelectric layer is electrically programmed using a deformable floating gate, resulting in analog conductance modulation based on the amplitude, number, and time interval of input magnetic pulses. This proposed extrasensory magnetoreceptive synapse serves as a synaptic compass for artificial intelligence, enabling obstacle-adaptive navigation and mapping of moving objects[130]. Inspired by hippocampal synapses, Lee et al. have created a dual-gate organic synaptic transistor platform featuring a photoconductive polymer semiconductor, a ferroelectric insulator P(VDF-TrFE), and a boric acid-functionalized extended gate electrode. This innovative design allows simultaneous detection of neurotransmitters, such as dopamine, and light. The developed synaptic transistor exhibits consolidated memory through repeated exposure to dopamine and polychromatic light, showcasing robustly modulated postsynaptic currents. This proof-of-concept for a hippocampus-synapse-mimicking organic neuromorphic system, integrating chemical sensors and photosensors, opens up new avenues for low-power organic artificial synapse multisensors and artificial synapses designed for light-induced memory consolidation. These advancements hold the potential to facilitate the development of machine interfaces[131].

### 3.4 FeFET based on two-dimensional materials

The physical limitations of silicon-based materials challenge the validity of Moore's Law. 2D materials are emerging as a promising candidate to overcome the bottleneck in CMOS scaling technology, holding significant potential for advancements in ultra-large-scale integration technology[132,133].

Conventional oxide ferroelectric materials face limitations related to thickness and complex preparation processes. As a result, researchers are increasingly favoring atomic-thin film ferroelectric materials, with a growing focus on 2D ferroelectric materials, marking a current research hotspot[134,135]. 2D materials are typically comprised of single or multiple atomic layers. Achieving ferroelectric switching in these materials generally requires at least two ions to form. The layers are held together by van der Waals forces, significantly weaker than ionic or covalent bonds. This characteristic allows for the mechanical exfoliation and transfer of thin films.

Notably, Guo and collaborators observed a bismuth layer with similarities to black phosphorus, demonstrating ordered charge transfer and regular atomic distortions between different sublattices[136,137]. In the bismuth monolayer, the weak and anisotropic sp orbital hybridization of bismuth atoms results in a buckled structure with broken inversion symmetry and charge redistribution within the unit cell. This unique configuration gives rise to in-plane electric polarization, extending the mechanism of ferroelectric properties to single-element materials.

As ferroelectric materials approach the two-dimensional limit, new phenomena emerge, and factors neglected in three-dimensional (3D) bulk materials become significant in two dimensions (2D), leading to differences between 2D and 3D ferroelectrics. At the nanoscale, the depolarization field of ferroelectric materials increases significantly, necessitating consideration of quantum confinement effects and tunneling effects. Additionally, surface-related effects are amplified due to the increased specific surface area in 2D ferroelectrics, leading to changes in crystal structures and chemical properties at the material's surface. These surface modifications can have both positive and negative impacts on ferroelectric properties[138]. For instance, while the depolarization field reduction at the nanoscale may decrease ferroelectric polarization, certain surface structures can enhance it[139]. Thus, modifying the surface structure of 2D ferroelectric materials can result in materials with robust ferroelectric properties.

2D ferroelectric materials have many advantages. Unlike traditional 3D ferroelectric materials, there are no dangling bonds on the surface of 2D vdW layered ferroelectric materials, which can effectively reduce the surface energy and facilitate the scaling and integration of devices. Due to the atomically thin structure of 2D materials, the in-plane metallic state and out-of-plane polarization are compatible, and ferroelectrics can exhibit two-dimensional superconductivity at low temperatures[42].

2D van der Waals (vdW) ferroelectrics exhibit enhanced robustness to the depolarization field. While the electric field generated by the depolarization field increases as the film thickness decreases, both polymer ferroelectrics and 2D vdW ferroelectrics maintain robust ferroelectricity even under extremely thin conditions. In contrast, conventional 3D ferroelectric thin films, while capable of maintaining ferroelectricity at the limit of thickness, require suitable substrate materials with minimal lattice mismatch for epitaxial growth. Layered materials, on the other hand, allow stacking of numerous layers with different structural properties, enabling the use of ferroelectric heterostructure devices independently of the substrate and providing broad tunability of ferroelectric properties[140]. The enhanced resistance of van der Waals (vdW) ferroelectrics to depolarizing fields can be attributed to several factors. Firstly, the ion displacement in 2D materials is larger than that in typical ferroelectric materials. For instance, in the single unit cell of $CuInP_2S_6$, the displacement of Cu ions can reach 1000 pm, whereas the ion displacement of Zr and Ti in PZT is 10 pm[141,142]. Or a semiconductor ferroelectric like α-$In_2Se_3$ has carriers that can move freely[39]. Second, 2D materials have smaller polarization values, which means they generate smaller depolarization fields[143,144]. Finally, the special properties of the material itself also play an important role in maintaining ferroelectricity, such as in-plane (OOP) and out-of-plane (IP) ferroelectricity in α-In$_2$Se$_3$[145]. In terms of optoelectronic devices, the potential of 2DvdW ferroelectrics is also very large, its band gap is relatively moderate, and its bandwidth can be adjusted by thickness[38]. Another interesting feature of 2D ferroelectrics is that their symmetry decreases as the dimension decreases. A typical example is SnS, the single layer is non-centrosymmetric and the block is centrosymmetric, and the number of layers is also It will change the symmetry of the material, and when the number of layers of the material is reduced to a few layers, it will produce odd and even effects[146], and the type of properties is also found in α-In$_2$Se$_3$[147]. 2D vdW ferroelectric materials can not only maintain ferroelectricity in single-layer or several-layer thickness, but also have rich material properties, including piezoelectricity, pyroelectricity, bulk photovoltaic effect, spontaneous spin polarization, valley pole etc[148]. The rich material properties and the tuning effect of ferroelectric polarization on the electronic properties of heterojunctions make 2D vdW ferroelectric materials have great application prospects in smart microdevices.

Research on van der Waals (vdW) ferroelectrics with 2D lattices has its roots in group IV semiconductor monochalcogenides dating back to the 1960s[35]. The discovery of the first 1mm-thick bulk CuInP2S6 occurred in 1994[149], but its thickness did not meet the requirements for electronic devices. The subsequent two decades saw limited breakthroughs in 2DvdW ferroelectrics. However, advancements in molecular beam epitaxy (MBE) technology and mechanical exfoliation significantly propelled the development of ultra-thin 2DvdW ferroelectric materials. In 2016, researchers reported stable ferroelectric polarization in CuInP2S6 with a thickness of 4 nm[40]. Since then, ferroelectric properties have been identified in numerous 2DvdW material systems, including In$_2$Se$_3$, CuCrP$_2$S$_6$, hexagonal boron nitride (h-BN), GeTe, MoTe$_2$, MoSe$_2$, SnSe, SnS, WTe$_2$, WSe$_2$, and WS$_2$, etc[41,146,150–158].

**Unique vdW ferroelectricity**

The ferroelectricity in both van der Waals (vdW) ferroelectrics and conventional ferroelectrics arises from asymmetric structures, leading to the spontaneous ordering of microscopic electric dipoles and the manifestation of macroscopic ferroelectric polarization. In 2DvdW materials, polarization is typically small, and interference from leakage currents complicates ferroelectricity measurements. Advanced techniques such as piezoresponse force microscopy (PFM), transmission electron microscopy (TEM), second harmonic generation (SHG), scanning tunneling microscopy (STM), and electrical transport measurements are employed to verify ferroelectricity. The verification process is particularly crucial due to the challenges posed by the small polarization and interference. To study the ferroelectric switching mechanism in 2D ferroelectric materials, researchers categorize them based on phase transition kinetics into displacement type (e.g., α-In2Se2), order-disorder type (e.g., CuInP2S6)[159], and mixed types. The ferroelectric switching mechanisms exhibit uniqueness across different vdW material systems, and the following provides a brief overview of some distinctive switching mechanisms. These insights aim to inspire improvements in device performance.

In$_2$Se$_3$ exhibits various crystal forms[160], with α-In2Se3 being the most stable phase characterized by a notable dipole-locking effect. This effect is rooted in the unique covalent bond configuration of In2Se3, which generates an electric field capable of simultaneously flipping in-plane and out-of-plane polarizations. Structurally, the reversal of in-plane and out-of-plane polarization occurs through the central Se atomic layer. The lateral movement of Se atoms alters the out-of-plane and in-plane interlayer spacing of In atoms, forming an intralayer covalent bond structure that produces both in-plane and out-of-plane dipoles. As the origin of these two dipoles is the same, the inversion of one dipole triggers the inversion of the other, leading to the "dipole locking effect.". Both in-plane and out-of-plane polarization reversals have been observed through SHG[161] and PFM[145], underscoring the importance of interlocking dipoles in maintaining ferroelectricity within the monoatomic layer and playing a crucial role in ferroelectric polarization switching[37]. In CuInP$_2$S$_6$, its unique properties stem from the special structure with a quadruple well potential and a movable Cu atom occupying the van der Waals (vdW) gap. One possible mechanism for CuInP$_2$S$_6$ polarization deflection involves the movement of Cu atoms approaching the material surface under the influence of an applied electric field. However, the specific behavior of Cu atoms on the material surface remains unclear. The movement of Cu atoms introduces novel ferroelectric phenomena, including negative piezoelectricity and inverse polarization switching, which were extensively discussed by Xue et al[37].

**2D vdW FeFET application**

Reservoir Computing (RC) is a type of recurrent neural network (RNN) characterized by low training and hardware costs. It finds applications in processing time series information, including waveform classification, speech recognition, and time series prediction. The system comprises two components: a reservoir pool that doesn't require training and an output layer that does. The reservoir layer can be implemented using a nonlinear device with short-range characteristics, exhibiting properties such as short-term memory and nonlinearity. By training only the output layer, the system substantially reduces training costs and enhances learning efficiency. Yang and Huang et al[162]. explored advanced RC hardware using α-In2Se3 materials. They initially developed a ferroelectric stackable RC based on α-In2Se3, enabling dynamic functions such as STP and LTP/LTD. A multi-layer RC system was constructed by linking the output of the primary RC to the input of the secondary RC(Figure 8a). This multi-layer RC demonstrated the capability to perform tasks like time series prediction and waveform classification, showcasing the high-dimensional mapping and computational potential of a deep RC architecture. Previous work by the team confirmed the system's viability for building reservoir computing hardware, and subsequent efforts utilized the electrical and optical dual-response mechanisms of ferroelectric synapses. These efforts aimed to achieve multi-mode and multi-scale signal processing, leveraging the short- and long-term plasticity of synapses based on ferroelectric conversion and charge capture/release processes. The synaptic device responds quickly after receiving the light signal, and the trap in the material produces slow attenuation and various short-term plasticity. The relaxation time of In$_2$Se$_3$ device decreases with the increase of applied voltage and light intensity, and the time interval of pulse signal changes the synaptic weight. Using the above features, the team uses it for multimode in-sensor pattern classification task and multiple-timescale reservoir computing task involving time-series prediction[45].

The development of two-dimensional material-based memory devices with optoelectronic responsivity in the short-wave infrared (SWIR) region, particularly for in-sensor RC at the optical communication band, presents a significant challenge. Zha et al.[163] have addressed this challenge by introducing an electronic/optoelectronic memory device based on a tellurium-containing two-dimensional van der Waals heterostructure. Operating in the SWIR region, this device can sense and process photon signals within the optical communication band. The heterostructure, combining ferroelectric $CuInP_2S_6$ with a tellurium channel, achieves both long-term and short-term memory effects through voltage pulses. Notably, it finds applications in an in-sensor reservoir computing system, enabling simultaneous sensing, decoding, and learning of information transmitted via optical fibers(Figure 8c).

Developing optoelectronic synapses is crucial for creating an advanced artificial visual system similar to human vision. However, achieving diverse functions of the biological visual neuromorphic system at the single-device level is challenging. A potential solution lies in 2D van der Waals (vdW) heterostructures. In this context, Guo et al[164]. propose a novel synaptic device utilizing a ferroelectric α-$In_2Se_3$/GaSe vdW heterojunction to replicate the entire biological visual system(Figure 8d). This device exhibits essential synaptic behaviors in response to both light and electrical stimuli, showcasing retina-like selectivity for light wavelengths. Beyond its optoelectronic synaptic functions, the device incorporates memory and logic functions reminiscent of the brain's visual cortex. The multifunctional synaptic device proposed in this study holds significant potential for processing intricate visual information.

Baek et al[165]. presents a van der Waals ferroelectrics heterostructure, $CuInP_2S_6$/α-$In_2Se_3$, integrated with a ferroelectric field-effect transistor (Fe-FET). The dipole coupling of CIPS and α-$In_2Se_3$ enhances polarization(Figure 8b), resulting in a 14.5 V memory window at VGS = ±10 V. Utilizing the Landau–Khalatnikov theory, the ferroelectric characteristics, hysteresis behaviors, and memory window formation are analyzed. The study also explores the potential applications of the ferroelectrics heterostructure in artificial synapses and hardware neural networks through training and inference simulations, offering insights into the integration of ferroelectric devices into advanced technologies. Soliman et al[166]. demonstrate precise control of ferroelectric polarization in $ReS_2$/hBN/$CuInP_2S_6$ van der Waals (vdW) FeFETs. Achieving an on/off ratio exceeding $10^7$ and a wide 7 V hysteresis window, these fully vdW-material FeFETs exhibit multiple remanent states with a lifetime exceeding $10^3$ s. Notably, optical control of $CuInP_2S_6$ polarization is achieved through photoexcitation. Operating in three synaptic modes—electrically stimulated, optically stimulated, and optically assisted—these FeFETs emulate essential synaptic functionalities.

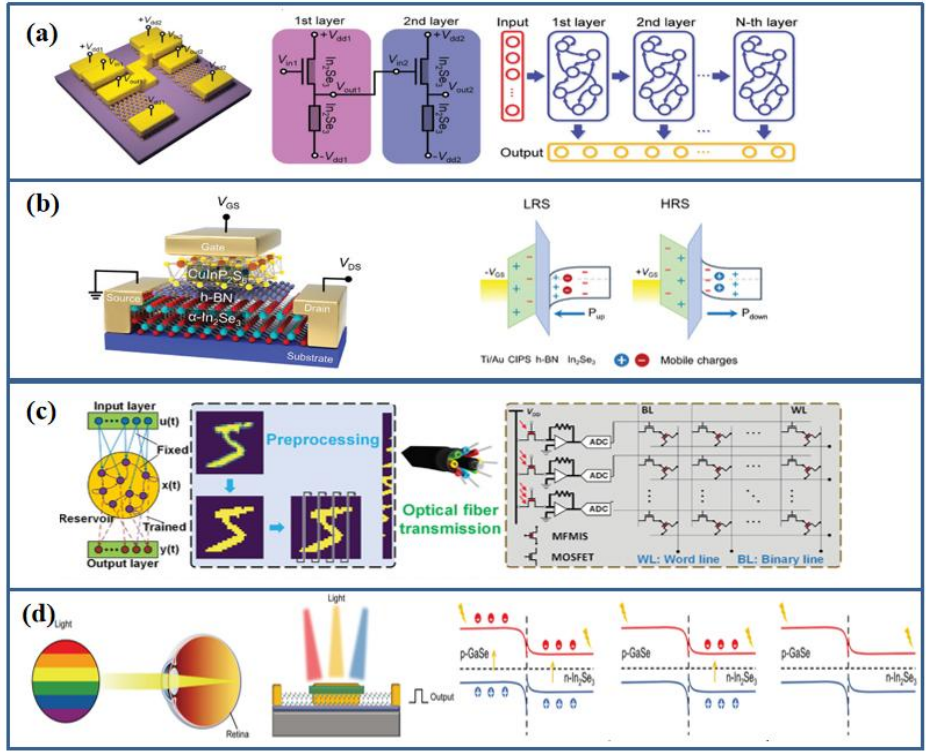

Figure 7 (a). Schematic diagram of the device structure, circuit principles, and architecture of a multi-layer RC system.(b). Band diagram of the α-In2Se3/h-BN/CIPS Fe-FET device in Pup state (left) and Pdown state (right).(c). Schematic diagram of the RC system used for classification on the MNIST dataset and confusion matrix for the MNIST test set classification.(d). Schematic diagram of OES (Optical-Electrical Stimulation) simulating the functioning of the retina, with an analysis of its response mechanism from a band diagram perspective on the right.

**Challenges and outlook**

Throughout the development experience of FeFET, people have explored deeply from the aspects of device structure optimization, material selection and new device principle. Extensive efforts have been dedicated to the enhancement of materials, the fabrication of individual devices, and the integration of large-scale arrays, with numerous challenges to address at each stage.

The perovskite oxide material system is currently the most thoroughly studied. Although it has unparalleled ferroelectricity, there are many reasons that limit its further application in neuromorphic devices. The first is the thickness of the material and poor bonding with Si. In order to perform large-scale integration, which means that large-scale circuits cannot be built to handle huge tasks, this problem has not been solved to this day. At the same time, this type of ferroelectric ceramics often contains lead elements that are toxic to the human body. The high volatility of lead elements will cause environmental pollution. People use alkali metals in the perovskite structure to replace lead to reduce the damage to nature. But it will reduce the ferroelectric performance to a certain extent. Halide perovskite materials can make up for the shortcomings of poor compatibility. The research on halide perovskite in neuromorphic devices has just begun, and this material is very susceptible to environmental influences. Its polarization mechanism is influenced by ion migration. Interfering and unclear. Therefore, people gradually turned their attention to finding other materials. Organic materials were subsequently used in the research of synaptic FeFETs. Compared with other devices, it has some obvious advantages. Reliability is a major advantage of organic synapses. Its polarization conversion is not as complicated as other materials. It will involve the migration of ions or defects, which reduces randomness to a certain extent. In addition, the flexibility and scalability of organic materials are the best among several material systems, and they have great potential in human-computer interaction devices. At present, there are some problems that need to be solved in organic synaptic devices. First, the residual polarization of PVDF and other materials is small, which is not conducive to the long-term data retention of synapses, which will cause irreversible errors and low accuracy in the calculation of synapses. Rate. Secondly, organic synaptic devices need to improve linearity and symmetry, which is crucial in some image information acquisition applications. If the polarization switching mechanism of polymer ferroelectric materials can be further explored, then the nonlinear problem of the device will be greatly improved. Hafnium-based materials are valued because of their high compatibility with the CMOS process. The use of such materials can maintain ferroelectricity even at extremely low thicknesses, contributing to a substantial increase in the memory window (MW) of synaptic devices. The physical properties of ferroelectricity observed in various forms of $HfO_2$-based ferroelectrics continue to pose challenges. The understanding of the strong ferroelectricity in $HfO_2$-based ferroelectrics remains unclear, hindering the further optimization, development, and practical applications of $HfO_2$-based ferroelectric devices. Numerous efforts have been made in this regard, such as delving into the oxygen vacancy mechanism and the preparation of more easily characterizable ferroelectric thin films. However, the current research progress is not yet systematically comprehensive. Therefore, the key issues and challenges of $HfO_2$-based ferroelectrics, spanning materials, device structures, and applications, focus on identifying the fundamental physical effects that govern the ferroelectric properties of thin films based on $HfO_2$. Two-dimensional materials, as strong contenders to perpetuate Moore's Law, face challenges in material preparation, interface optimization, repeatability, and stability. Addressing these issues is crucial for the advancement of 2D materials. Furthermore, the exploration of alternative device structures and process technologies is indispensable in achieving success. The positive outcomes of these endeavors anticipate the pivotal role of van der Waals Field-Effect Transistors (vdWs FeFETs) in shaping the next generation of high-performance semiconductor devices. In addition, a challenge for two-dimensional materials is the difficulty of achieving large-scale growth, which is crucial for achieving extensive integration. The key to preparing wafer-level devices lies in obtaining large-area, high-quality 2D van der Waals (vdWs) ferroelectric thin films. Traditional mechanical peeling methods struggle to guarantee both the area and quality of samples. Hence, it is necessary to explore chip-level growth methods suitable for 2D vdWs ferroelectric films, ensuring that high-quality film samples meet the requirements of device manufacturing.

In summary, synaptic transistor devices based on ferroelectric materials offer advantages such as simple structure, low power consumption, high on/off ratio, and fast response speed, making them particularly suitable for neuromorphic computing applications. Future research should focus on several key aspects. First, attention should be directed towards understanding the ferroelectric conversion mechanism and internal physics of the materials themselves. Suitable combinations of ferroelectric semiconductor layers and ferroelectric insulator layers need to be identified for device integration in various application scenarios. Second, optimization and evaluation of synaptic device performance and electrical properties are crucial. Issues such as leakage current and uniformity stability in ferroelectric transistors, as well as linearity and symmetry problems in synaptic behavior, need to be addressed. Third, in terms of device integration, there is a scarcity of reports on large-scale ferroelectric arrays, and their operational characteristics and reliability still require validation. Crosstalk and noise issues in large-scale arrays also need careful consideration. For large-scale arrays applied to neuromorphic computing, the development and utilization of pulse-enhancement/inhibition with the same pulse for simultaneous learning behavior and signal transmission are essential. Finally, in the application research of ferroelectric synapses, current algorithms exhibit lower accuracy compared to mature deep neural networks (DNNs) and involve substantial learning costs. Power consumption remains a significant challenge for practical applications of ferroelectric arrays. On one hand, the need for read-before-write operations in ferroelectric transistors increases circuit complexity and energy consumption. On the other hand, a suitable low-power Spiking Neural Network (SNN) architecture has yet to be identified. The practical application of ferroelectric synapses is still in the early stages of principle verification, such as simple digit recognition in image processing, and requires further enhancement of its capabilities for handling more complex tasks.

参考文献

1. Kumar, S., Wang, X., Strachan, J. P., Yang, Y. & Lu, W. D. Dynamical memristors for higher-complexity neuromorphic computing. *Nat Rev Mater* **7**, 575–591 (2022).

2. He, Y. *et al.* Recent Progress on Emerging Transistor-Based Neuromorphic Devices. *Advanced Intelligent Systems* **3**, 2000210 (2021).

3. Yan, S. *et al.* Recent progress in ferroelectric synapses and their applications. *Sci. China Mater.* **66**, 877–894 (2023).

4. Merolla, P. A. *et al.* A million spiking-neuron integrated circuit with a scalable communication network and interface. *Science* **345**, 668–673 (2014).

5. Sarkar, D. *et al.* Mimicking Biological Synaptic Functionality with an Indium Phosphide Synaptic Device on Silicon for Scalable Neuromorphic Computing. *ACS Nano* **12**, 1656–1663 (2018).

6. Chen, S., Zhang, T., Tappertzhofen, S., Yang, Y. & Valov, I. Electrochemical Memristor-Based Artificial Neurons and Synapses – fundamentals, applications, and challenges. *Advanced Materials* 2301924 (2023) doi:10.1002/adma.202301924.

7. Wang, Z. *et al.* Fully memristive neural networks for pattern classification with unsupervised learning. *Nat Electron* **1**, 137–145 (2018).

8. Cai, Y. *et al.* Molecular ferroelectric/semiconductor interfacial memristors for artificial synapses. *npj Flex Electron* **6**, 16 (2022).

9. Ling, S., Zhang, C., Ma, C., Li, Y. & Zhang, Q. Emerging MXene-Based Memristors for In-Memory, Neuromorphic Computing, and Logic Operation. *Adv Funct Materials* **33**, 2208320 (2023).

10. Liu, H. *et al.* Artificial Neuronal Devices Based on Emerging Materials: Neuronal

Dynamics and Applications. *Advanced Materials* 2205047 (2023) doi:10.1002/adma.202205047.

11. Strukov, D. B., Snider, G. S., Stewart, D. R. & Williams, R. S. The missing memristor found. *Nature* **453**, 80–83 (2008).

12. Fuller, E. J. *et al.* Parallel programming of an ionic floating-gate memory array for scalable neuromorphic computing. (2019).

13. Niu, X., Tian, B., Zhu, Q., Dkhil, B. & Duan, C. Ferroelectric polymers for neuromorphic computing. *Applied Physics Reviews* **9**, 021309 (2022).

14. Wang, J. *et al.* Logic and in-memory computing achieved in a single ferroelectric semiconductor transistor. *Science Bulletin* **66**, 2288–2296 (2021).

15. Mikolajick, T., Park, M. H., Begon-Lours, L. & Slesazeck, S. From Ferroelectric Material Optimization to Neuromorphic Devices. *Advanced Materials* 2206042 (2023) doi:10.1002/adma.202206042.

16. Jiao, H. *et al.* Ferroelectric field effect transistors for electronics and optoelectronics. *Applied Physics Reviews* **10**, 011310 (2023).

17. Wang, Z. *et al.* Resistive switching materials for information processing. *Nat Rev Mater* **5**, 173–195 (2020).

18. Zhu, J., Zhang, T., Yang, Y. & Huang, R. A comprehensive review on emerging artificial neuromorphic devices. *Applied Physics Reviews* **7**, 011312 (2020).

19. Dai, S. *et al.* Recent Advances in Transistor-Based Artificial Synapses. *Adv. Funct. Mater.* **29**, 1903700 (2019).

20. Yu, H. *et al.* Evolution of Bio-Inspired Artificial Synapses: Materials, Structures, and

Mechanisms. *Small* **17**, 2000041 (2021).

21. Li, X. *et al.* Flexible Artificial Synapses Based on Field Effect Transistors: From Materials, Mechanics towards Applications. *Advanced Intelligent Systems* **4**, 2200015 (2022).

22. Lee, Y., Park, H.-L., Kim, Y. & Lee, T.-W. Organic electronic synapses with low energy consumption. *Joule* **5**, 794–810 (2021).

23. Valasek, J. Piezo-Electric and Allied Phenomena in Rochelle Salt. *Phys. Rev.* **17**, 475–481 (1921).

24. Haeni, J. H. *et al.* Room-temperature ferroelectricity in strained SrTiO3. **430**, (2004).

25. Xu, R. *et al.* Strain-induced room-temperature ferroelectricity in SrTiO3 membranes. *Nat Commun* **11**, 3141 (2020).

26. Zhao, X., Menzel, S., Polian, I., Schmidt, H. & Du, N. Review on Resistive Switching Devices Based on Multiferroic BiFeO3. *Nanomaterials* **13**, 1325 (2023).

27. Liao, W.-Q. *et al.* A lead-halide perovskite molecular ferroelectric semiconductor. *Nat Commun* **6**, 7338 (2015).

28. Su, L. *et al.* Flexible, Fatigue-Free, and Large-Scale $Bi_{3.25}La_{0.75}Ti_3O_{12}$ Ferroelectric Memories. *ACS Appl. Mater. Interfaces* **10**, 21428–21433 (2018).

29. Jeong, Y. *et al.* Low Voltage and Ferroelectric 2D Electron Devices Using Lead-Free $Ba_xSr_{1-x}TiO_3$ and $MoS_2$ Channel. *Adv. Funct. Mater.* **30**, 1908210 (2020).

30. Mulaosmanovic, H. *et al.* Ferroelectric field-effect transistors based on $HfO_2$: a review. *Nanotechnology* **32**, 502002 (2021).

31. Davis, G. T., McKinney, J. E., Broadhurst, M. G. & Roth, S. C. Electric-field-induced phase changes in poly(vinylidene fluoride). *Journal of Applied Physics* **49**, 4998–5002 (1978).

32. Furukawa, T., Date, M., Fukada, E., Tajitsu, Y. & Chiba, A. Ferroelectric Behavior in the Copolymer of Vinylidenefluoride and Trifluoroethylene. *Jpn. J. Appl. Phys.* **19**, L109–L112 (1980).

33. Hiroshi Ishiwara, H. I. Proposal of Adaptive-Learning Neuron Circuits with Ferroelectric Analog-Memory Weights. *Jpn. J. Appl. Phys.* **32**, 442 (1993).

34. Schroeder, U., Park, M. H., Mikolajick, T. & Hwang, C. S. The fundamentals and applications of ferroelectric HfO2. *Nat Rev Mater* **7**, 653–669 (2022).

35. Wang, C., You, L., Cobden, D. & Wang, J. Towards two-dimensional van der Waals ferroelectrics. *Nat. Mater.* **22**, 542–552 (2023).

36. Cui, C., Xue, F., Hu, W.-J. & Li, L.-J. Two-dimensional materials with piezoelectric and ferroelectric functionalities. *npj 2D Mater Appl* **2**, 18 (2018).

37. Xue, F., He, J.-H. & Zhang, X. Emerging van der Waals ferroelectrics: Unique properties and novel devices. *Applied Physics Reviews* **8**, 021316 (2021).

38. Zhang, D., Schoenherr, P., Sharma, P. & Seidel, J. Ferroelectric order in van der Waals layered materials. *Nat Rev Mater* **8**, 25–40 (2022).

39. Ding, W. *et al.* Prediction of intrinsic two-dimensional ferroelectrics in In2Se3 and other III2-VI3 van der Waals materials. *Nat Commun* **8**, 14956 (2017).

40. Liu, F. *et al.* Room-temperature ferroelectricity in CuInP2S6 ultrathin flakes. *Nat Commun* **7**, 12357 (2016).

41. Yuan, S. *et al.* Room-temperature ferroelectricity in MoTe2 down to the atomic monolayer limit. *Nat Commun* **10**, 1775 (2019).

42. Jindal, A. *et al.* Coupled ferroelectricity and superconductivity in bilayer Td-MoTe2.

*Nature* **613**, 48–52 (2023).

43. Wang, X. *et al.* Interfacial ferroelectricity in rhombohedral-stacked bilayer transition metal dichalcogenides. *Nat. Nanotechnol.* **17**, 367–371 (2022).

44. Andersen, T. I. *et al.* Excitons in a reconstructed moiré potential in twisted WSe2/WSe2 homobilayers. *Nat. Mater.* **20**, 480–487 (2021).

45. Liu, K. *et al.* An optoelectronic synapse based on α-In2Se3 with controllable temporal dynamics for multimode and multiscale reservoir computing. *Nat Electron* **5**, 761–773 (2022).

46. Xue, F. *et al.* Optoelectronic Ferroelectric Domain-Wall Memories Made from a Single Van Der Waals Ferroelectric. *Adv. Funct. Mater.* **30**, 2004206 (2020).

47. Zheng, W. *et al.* Emerging Halide Perovskite Ferroelectrics. *Advanced Materials* **35**, 2205410 (2023).

48. Cai, Y. *et al.* van der Waals Ferroelectric Halide Perovskite Artificial Synapse. *Phys. Rev. Applied* **18**, 014014 (2022).

49. Yao, Y. *et al.* Exploring a Fatigue-Free Layered Hybrid Perovskite Ferroelectric for Photovoltaic Non-Volatile Memories. *Angewandte Chemie International Edition* **60**, 10598–10602 (2021).

50. Sung-Min Yoon, Tokumitsu, E. & Ishiwara, H. An electrically modifiable synapse array composed of metal-ferroelectric-semiconductor (MFS) FET's using SrBi/sub 2/Ta/sub 2/O/sub 9/ thin films. *IEEE Electron Device Lett.* **20**, 229–231 (1999).

51. Sung-Min Yoon, S.-M. Y., Eisuke Tokumitsu, E. T. & Hiroshi Ishiwara, H. I. Improvement of Memory Retention Characteristics in Ferroelectric Neuron Circuits Using a Pt/SrBi$_2$Ta$_2$

O$_9$/Pt/Ti/SiO$_2$/Si Structure-Field Effect Transistor as a Synapse Device. *Jpn. J. Appl. Phys.* **39**, 2119 (2000).

52. Yang, A. J. *et al.* Two-Dimensional Layered Materials Meet Perovskite Oxides: A Combination for High-Performance Electronic Devices. *ACS Nano* acsnano.3c00429 (2023) doi:10.1021/acsnano.3c00429.

53. Du, J. *et al.* A robust neuromorphic vision sensor with optical control of ferroelectric switching. *Nano Energy* **89**, 106439 (2021).

54. Puebla, S. *et al.* Combining Freestanding Ferroelectric Perovskite Oxides with Two-Dimensional Semiconductors for High Performance Transistors. *Nano Lett.* **22**, 7457–7466 (2022).

55. Liu, X. *et al.* Charge–Ferroelectric Transition in Ultrathin Na$_{0.5}$Bi$_{4.5}$Ti$_4$O$_{15}$ Flakes Probed via a Dual-Gated Full van der Waals Transistor. *Advanced Materials* **32**, 2004813 (2020).

56. Luo, Z.-D. *et al.* Artificial Optoelectronic Synapses Based on Ferroelectric Field-Effect Enabled 2D Transition Metal Dichalcogenide Memristive Transistors. *ACS Nano* **14**, 746–754 (2020).

57. Böscke, T. S., Müller, J., Bräuhaus, D., Schröder, U. & Böttger, U. Ferroelectricity in hafnium oxide thin films. *Applied Physics Letters* **99**, 102903 (2011).

58. Starschich, S., Schenk, T., Schroeder, U. & Boettger, U. Ferroelectric and piezoelectric properties of Hf1-xZrxO2 and pure ZrO2 films. *Applied Physics Letters* **110**, 182905 (2017).

59. Silva, J. P. B. *et al.* Wake-up Free Ferroelectric Rhombohedral Phase in Epitaxially Strained ZrO$_2$ Thin Films. *ACS Appl. Mater. Interfaces* **13**, 51383–51392 (2021).

60. Khan, A. I., Keshavarzi, A. & Datta, S. The future of ferroelectric field-effect transistor technology. *Nat Electron* **3**, 588–597 (2020).

61. Müller, J. *et al.* Ferroelectricity in HfO2 enables nonvolatile data storage in 28 nm HKMG. in *2012 Symposium on VLSI Technology (VLSIT)* 25–26 (2012). doi:10.1109/VLSIT.2012.6242443.

62. Dünkel, S. *et al.* A FeFET based super-low-power ultra-fast embedded NVM technology for 22nm FDSOI and beyond. in *2017 IEEE International Electron Devices Meeting (IEDM)* 19.7.1-19.7.4 (2017). doi:10.1109/IEDM.2017.8268425.

63. Nukala, P. *et al.* Reversible oxygen migration and phase transitions in hafnia-based ferroelectric devices. *Science* **372**, 630–635 (2021).

64. Lee, H.-J. *et al.* Scale-free ferroelectricity induced by flat phonon bands in $HfO_2$. *Science* **369**, 1343–1347 (2020).

65. Lee, D. H. *et al.* Neuromorphic devices based on fluorite-structured ferroelectrics. *InfoMat* **4**, (2022).

66. Cheema, S. S. *et al.* Enhanced ferroelectricity in ultrathin films grown directly on silicon. *Nature* **580**, 478–482 (2020).

67. Noheda, B. & Íñiguez, J. A key piece of the ferroelectric hafnia puzzle. *Science* **369**, 1300–1301 (2020).

68. Ohtaka, O. *et al.* Phase Relations and Volume Changes of Hafnia under High Pressure and High Temperature. *Journal of the American Ceramic Society* **84**, 1369–1373 (2004).

69. Ohtaka, O. *et al.* Phase relations and equations of state of $ZrO_2$ under high temperature and high pressure. *Phys. Rev. B* **63**, 174108 (2001).

70. Xu, X. *et al.* Kinetically stabilized ferroelectricity in bulk single-crystalline HfO2:Y. *Nat. Mater.* **20**, 826–832 (2021).

71. Hsain, H. A. *et al.* Many routes to ferroelectric $HfO_2$: A review of current deposition methods. *Journal of Vacuum Science & Technology A* **40**, 010803 (2022).

72. Zhong, H. *et al.* Large-Scale $Hf_{0.5}Zr_{0.5}O_2$ Membranes with Robust Ferroelectricity. *Advanced Materials* **34**, 2109889 (2022).

73. Li, T. *et al.* Interface control of tetragonal ferroelectric phase in ultrathin Si-doped HfO2 epitaxial films. *Acta Materialia* **207**, 116696 (2021).

74. Müller, J. *et al.* Ferroelectricity in Simple Binary $ZrO_2$ and $HfO_2$. *Nano Lett.* **12**, 4318–4323 (2012).

75. Tang, L. *et al.* Regulating crystal structure and ferroelectricity in Sr doped HfO2 thin films fabricated by metallo-organic decomposition. *Ceramics International* **45**, 3140–3147 (2019).

76. Shiraishi, T. *et al.* Formation of the orthorhombic phase in CeO2-HfO2 solid solution epitaxial thin films and their ferroelectric properties. *Applied Physics Letters* **114**, 232902 (2019).

77. Yao, Y. *et al.* Experimental evidence of ferroelectricity in calcium doped hafnium oxide thin films. *Journal of Applied Physics* **126**, 154103 (2019).

78. Starschich, S. & Boettger, U. An extensive study of the influence of dopants on the ferroelectric properties of $HfO_2$. *J. Mater. Chem. C* **5**, 333–338 (2017).

79. Shiraishi, T. *et al.* Fabrication of ferroelectric Fe doped $HfO_2$ epitaxial thin films by ion-beam sputtering method and their characterization. *Jpn. J. Appl. Phys.* **57**, 11UF02 (2018).

80. Liu, H. *et al.* Structural and ferroelectric properties of Pr doped HfO2 thin films fabricated by chemical solution method. *J Mater Sci: Mater Electron* **30**, 5771–5779 (2019).

81. Tromm, T. C. U. *et al.* Ferroelectricity in Lu doped HfO2 layers. *Applied Physics Letters* **111**, 142904 (2017).

82. Cao, R. *et al.* Compact artificial neuron based on anti-ferroelectric transistor. *Nat Commun* **13**, 7018 (2022).

83. Kim, K. D. *et al.* Ferroelectricity in undoped-$HfO_2$ thin films induced by deposition temperature control during atomic layer deposition. *J. Mater. Chem. C* **4**, 6864–6872 (2016).

84. Hyuk Park, M. *et al.* Effect of forming gas annealing on the ferroelectric properties of Hf0.5Zr0.5O2 thin films with and without Pt electrodes. *Applied Physics Letters* **102**, 112914 (2013).

85. Luo, J.-D. *et al.* Correlation between ferroelectricity and nitrogen incorporation of undoped hafnium dioxide thin films. *Vacuum* **176**, 109317 (2020).

86. Park, M. H. *et al.* Understanding the formation of the metastable ferroelectric phase in hafnia–zirconia solid solution thin films. *Nanoscale* **10**, 716–725 (2018).

87. Park, M. H., Lee, Y. H. & Hwang, C. S. Understanding ferroelectric phase formation in doped $HfO_2$ thin films based on classical nucleation theory. *Nanoscale* **11**, 19477–19487 (2019).

88. Chen, H. *et al.* $HfO_2$-based ferroelectrics: From enhancing performance, material design, to applications. *Applied Physics Reviews* **9**, 011307 (2022).

89. Yurchuk, E. *et al.* Impact of layer thickness on the ferroelectric behaviour of silicon doped hafnium oxide thin films. *Thin Solid Films* **533**, 88–92 (2013).

90. Hyuk Park, M. *et al.* Evolution of phases and ferroelectric properties of thin Hf0.5Zr0.5O2 films according to the thickness and annealing temperature. *Applied Physics Letters* **102**, 242905 (2013).

91. Kim, S. J. *et al.* Large ferroelectric polarization of TiN/Hf0.5Zr0.5O2/TiN capacitors due to stress-induced crystallization at low thermal budget. *Applied Physics Letters* **111**, 242901 (2017).

92. Shiraishi, T. *et al.* Impact of mechanical stress on ferroelectricity in (Hf0.5Zr0.5)O2 thin films. *Applied Physics Letters* **108**, 262904 (2016).

93. Kim, M.-K. & Lee, J.-S. Ferroelectric Analog Synaptic Transistors. *Nano Lett.* **19**, 2044–2050 (2019).

94. Kim, D. *et al.* Analog Synaptic Transistor with Al-Doped $HfO_2$ Ferroelectric Thin Film. *ACS Appl. Mater. Interfaces* **13**, 52743–52753 (2021).

95. Yang, X. *et al.* A self-powered artificial retina perception system for image preprocessing based on photovoltaic devices and memristive arrays. *Nano Energy* **78**, 105246 (2020).

96. Seo, S. *et al.* An Optogenetics-Inspired Flexible van der Waals Optoelectronic Synapse and its Application to a Convolutional Neural Network. *Advanced Materials* **33**, 2102980 (2021).

97. Kim, M.-K., Kim, I.-J. & Lee, J.-S. CMOS-compatible compute-in-memory accelerators based on integrated ferroelectric synaptic arrays for convolution neural networks.

SCIENCE ADVANCES **8**, eabm8537 (2022).

98. Kim, M. & Lee, J. Synergistic Improvement of Long-Term Plasticity in Photonic Synapses Using Ferroelectric Polarization in Hafnia-Based Oxide-Semiconductor Transistors. *Adv. Mater.* **32**, 1907826 (2020).

99. Jeon, H. *et al.* Hysteresis Modulation on Van der Waals-Based Ferroelectric Field-Effect Transistor by Interfacial Passivation Technique and Its Application in Optic Neural Networks. *Small* **16**, 2004371 (2020).

100. Joh, H. *et al.* Flexible Ferroelectric Hafnia-Based Synaptic Transistor by Focused-Microwave Annealing. *ACS Appl. Mater. Interfaces* **14**, 1326–1333 (2022).

101. Xia, F. *et al.* Carbon Nanotube-Based Flexible Ferroelectric Synaptic Transistors for Neuromorphic Computing. *ACS Appl. Mater. Interfaces* **14**, 30124–30132 (2022).

102. Anwar, S. *et al.* Solution-processed transparent ferroelectric nylon thin films. *Sci. Adv.* **5**, eaav3489 (2019).

103. Horiuchi, S. *et al.* Above-room-temperature ferroelectricity in a single-component molecular crystal. *Nature* **463**, 789–792 (2010).

104. Funatsu, Y., Sonoda, A. & Funahashi, M. Ferroelectric liquid-crystalline semiconductors based on a phenylterthiophene skeleton: effect of the introduction of oligosiloxane moieties and photovoltaic effect. *J. Mater. Chem. C* **3**, 1982–1993 (2015).

105. Urbanaviciute, I. *et al.* Negative piezoelectric effect in an organic supramolecular ferroelectric. *Mater. Horiz.* **6**, 1688–1698 (2019).

106. Kawai, H. The Piezoelectricity of Poly (vinylidene Fluoride). *Jpn. J. Appl. Phys.* **8**, 975 (1969).

107. Li, M. *et al.* Revisiting the δ-phase of poly(vinylidene fluoride) for solution-processed ferroelectric thin films. *Nature Mater* **12**, 433–438 (2013).

108. Saxena, P. & Shukla, P. A comprehensive review on fundamental properties and applications of poly(vinylidene fluoride) (PVDF). *Adv Compos Hybrid Mater* **4**, 8–26 (2021).

109. Zhu, H., Yamamoto, S., Matsui, J., Miyashita, T. & Mitsuishi, M. Ferroelectricity of poly(vinylidene fluoride) homopolymer Langmuir–Blodgett nanofilms. *J. Mater. Chem. C* **2**, 6727–6731 (2014).

110. Zhu, H., Fu, C. & Mitsuishi, M. Organic ferroelectric field-effect transistor memories with POLY(VINYLIDENE FLUORIDE) gate insulators and conjugated semiconductor channels: a review. *Polym Int* **70**, 404–413 (2021).

111. Ni, Y., Wang, Y. & Xu, W. Recent Process of Flexible Transistor-Structured Memory. *Small* **17**, 1905332 (2021).

112. Xu, T., Xiang, L., Xu, M., Xie, W. & Wang, W. Excellent low-voltage operating flexible ferroelectric organic transistor nonvolatile memory with a sandwiching ultrathin ferroelectric film. *Sci Rep* **7**, 8890 (2017).

113. Gao, S., Yi, X., Shang, J., Liu, G. & Li, R.-W. Organic and hybrid resistive switching materials and devices. *Chem. Soc. Rev.* **48**, 1531–1565 (2019).

114. Xu, M., Guo, S., Xu, T., Xie, W. & Wang, W. Low-voltage programmable/erasable high performance flexible organic transistor nonvolatile memory based on a tetratetracontane passivated ferroelectric terpolymer. *Organic Electronics* **64**, 62–70 (2019).

115. Chen, X., Han, X. & Shen, Q.-D. PVDF-Based Ferroelectric Polymers in Modern Flexible

Electronics. *Adv. Electron. Mater.* **3**, 1600460 (2017).

116. Martins, P., Lopes, A. C. & Lanceros-Mendez, S. Electroactive phases of poly(vinylidene fluoride): Determination, processing and applications. *Progress in Polymer Science* **39**, 683–706 (2014).

117. Song, W.-Z. *et al.* Single electrode piezoelectric nanogenerator for intelligent passive daytime radiative cooling. *Nano Energy* **82**, 105695 (2021).

118. Dawson, N. M., Atencio, P. M. & Malloy, K. J. Facile deposition of high quality ferroelectric poly(vinylidene fluoride) thin films by thermally modulated spin coating. *J. Polym. Sci. Part B: Polym. Phys.* **55**, 221–227 (2017).

119. Park, K.-W. *et al.* Humidity effect of domain wall roughening behavior in ferroelectric copolymer thin films. *Nanotechnology* **25**, 355703 (2014).

120. Qian, X., Chen, X., Zhu, L. & Zhang, Q. M. Fluoropolymer ferroelectrics: Multifunctional platform for polar-structured energy conversion. *Science* **380**, eadg0902 (2023).

121. Dang, Z., Guo, F., Wu, Z., Jin, K. & Hao, J. Interface Engineering and Device Applications of 2D Ultrathin Film/Ferroelectric Copolymer P(VDF-TrFE). *Advanced Physics Research* **2**, 2200038 (2023).

122. Xie, P. *et al.* Ferroelectric P(VDF-TrFE) wrapped InGaAs nanowires for ultralow-power artificial synapses. *Nano Energy* **91**, 106654 (2022).

123. Dang, Z. *et al.* Black Phosphorus/Ferroelectric P(VDF-TrFE) Field-Effect Transistors with High Mobility for Energy-Efficient Artificial Synapse in High-Accuracy Neuromorphic Computing. *Nano Lett.* acs.nanolett.3c01687 (2023) doi:10.1021/acs.nanolett.3c01687.

124. Yan, M. *et al.* Ferroelectric Synaptic Transistor Network for Associative Memory. *Adv.*

*Electron. Mater.* **7**, 2001276 (2021).

125. Lee, Y. R., Trung, T. Q., Hwang, B.-U. & Lee, N.-E. A flexible artificial intrinsic-synaptic tactile sensory organ. *Nat Commun* **11**, 2753 (2020).

126. Lee, K. *et al.* Artificially Intelligent Tactile Ferroelectric Skin. *Adv. Sci.* **7**, 2001662 (2020).

127. Yu, R. *et al.* Programmable ferroelectric bionic vision hardware with selective attention for high-precision image classification. *Nat Commun* **13**, 7019 (2022).

128. Li, Q. *et al.* Ultralow Power Wearable Organic Ferroelectric Device for Optoelectronic Neuromorphic Computing. *Nano Lett.* **22**, 6435–6443 (2022).

129. Jiang, Y. *et al.* Asymmetric Ferroelectric-Gated Two-Dimensional Transistor Integrating Self-Rectifying Photoelectric Memory and Artificial Synapse. *ACS Nano* **16**, 11218–11226 (2022).

130. Kim, Y. *et al.* Bird-Inspired Self-Navigating Artificial Synaptic Compass. *ACS Nano* **15**, 20116–20126 (2021).

131. Lee, H. R., Lee, D. & Oh, J. H. A Hippocampus-Inspired Dual-Gated Organic Artificial Synapse for Simultaneous Sensing of a Neurotransmitter and Light. *Advanced Materials* **33**, 2100119 (2021).

132. Das, S. *et al.* Transistors based on two-dimensional materials for future integrated circuits. *Nat Electron* **4**, 786–799 (2021).

133. Wu, F. *et al.* Vertical MoS2 transistors with sub-1-nm gate lengths. *Nature* **603**, 259–264 (2022).

134. Han, W. *et al.* Phase-controllable large-area two-dimensional In2Se3 and ferroelectric heterophase junction. *Nat. Nanotechnol.* **18**, 55–63 (2023).

135. Weston, A. *et al.* Interfacial ferroelectricity in marginally twisted 2D semiconductors. *Nat. Nanotechnol.* **17**, 390–395 (2022).

136. Gou, J. *et al.* The effect of moiré superstructures on topological edge states in twisted bismuthene homojunctions. *Sci. Adv.* **6**, eaba2773 (2020).

137. Gou, J. *et al.* Two-dimensional ferroelectricity in a single-element bismuth monolayer. *Nature* **617**, 67–72 (2023).

138. Qi, L., Ruan, S. & Zeng, Y. Review on Recent Developments in 2D Ferroelectrics: Theories and Applications. *Adv. Mater.* **33**, 2005098 (2021).

139. Chang, K. *et al.* Discovery of robust in-plane ferroelectricity in atomic-thick SnTe. *Science* **353**, 274–278 (2016).

140. Han, M. *et al.* Continuously tunable ferroelectric domain width down to the single-atomic limit in bismuth tellurite. *Nat Commun* **13**, 5903 (2022).

141. Jia, C.-L. *et al.* Atomic-scale study of electric dipoles near charged and uncharged domain walls in ferroelectric films. *Nature Mater* **7**, 57–61 (2008).

142. Neumayer, S. M. *et al.* Alignment of Polarization against an Electric Field in van der Waals Ferroelectrics. *Phys. Rev. Applied* **13**, 064063 (2020).

143. Rodriguez, J. R. *et al.* Electric field induced metallic behavior in thin crystals of ferroelectric $\alpha$-In2Se3. *Applied Physics Letters* **117**, 052901 (2020).

144. Sharma, P. *et al.* A room-temperature ferroelectric semimetal. *Sci. Adv.* **5**, eaax5080 (2019).

145. Xue, F. *et al.* Room-Temperature Ferroelectricity in Hexagonally Layered α-In2Se3 Nanoflakes down to the Monolayer Limit. *Advanced Functional Materials* **28**, 1803738

(2018).

146. Higashitarumizu, N. *et al.* Purely in-plane ferroelectricity in monolayer SnS at room temperature. *Nat Commun* **11**, 2428 (2020).

147. Lv, B. *et al.* Layer-dependent ferroelectricity in 2H-stacked few-layer α-In$_2$Se$_3$. *Mater. Horiz.* **8**, 1472–1480 (2021).

148. Li, Y. *et al.* Enhanced bulk photovoltaic effect in two-dimensional ferroelectric CuInP2S6. *Nat Commun* **12**, 5896 (2021).

149. Simon, A., Ravez, J., Maisonneuve, V., Payen, C. & Cajipe, V. B. Paraelectric-Ferroelectric Transition in the Lamellar Thiophosphate CuInP2S6. *Chem. Mater.* **6**, 1575–1580 (1994).

150. Varotto, S. *et al.* Room-temperature ferroelectric switching of spin-to-charge conversion in germanium telluride. *Nat Electron* **4**, 740–747 (2021).

151. Zhou, Y. *et al.* Out-of-Plane Piezoelectricity and Ferroelectricity in Layered α-In$_2$Se$_3$ Nanoflakes. *Nano Lett.* **17**, 5508–5513 (2017).

152. Zheng, Z. *et al.* Unconventional ferroelectricity in moiré heterostructures. *Nature* **588**, 71–76 (2020).

153. Cho, K. *et al.* Tunable Ferroelectricity in Van der Waals Layered Antiferroelectric CuCrP$_2$S$_6$. *Adv Funct Materials* **32**, 2204214 (2022).

154. Sung, J. *et al.* Broken mirror symmetry in excitonic response of reconstructed domains in twisted MoSe2/MoSe2 bilayers. *Nat. Nanotechnol.* **15**, 750–754 (2020).

155. Du, R. *et al.* Two-dimensional multiferroic material of metallic p-doped SnSe. *Nat Commun* **13**, 6130 (2022).

156. Fei, Z. *et al.* Ferroelectric switching of a two-dimensional metal. *Nature* **560**, 336–339

(2018).

157. Liu, Y., Liu, S., Li, B., Yoo, W. J. & Hone, J. Identifying the Transition Order in an Artificial Ferroelectric van der Waals Heterostructure. *Nano Lett.* **22**, 1265–1269 (2022).

158. Rogée, L. *et al.* Ferroelectricity in untwisted heterobilayers of transition metal dichalcogenides. *Science* **376**, 973–978 (2022).

159. Zhou, S. *et al.* Van der Waals Layered Ferroelectric CuInP2S6: Physical Properties and Device Applications. Preprint at http://arxiv.org/abs/2009.02097 (2020).

160. Tan, C. K. Y., Fu, W. & Loh, K. P. Polymorphism and Ferroelectricity in Indium(III) Selenide. *Chem. Rev.* **123**, 8701–8717 (2023).

161. Xiao, J. *et al.* Intrinsic Two-Dimensional Ferroelectricity with Dipole Locking. *Phys. Rev. Lett.* **120**, 227601 (2018).

162. Liu, K. *et al.* Multilayer Reservoir Computing Based on Ferroelectric α-In$_2$Se$_3$ for Hierarchical Information Processing. *Advanced Materials* **34**, 2108826 (2022).

163. Zha, J. *et al.* Electronic/Optoelectronic Memory Device Enabled by Tellurium-based 2D van der Waals Heterostructure for in-Sensor Reservoir Computing at the Optical Communication Band. *Advanced Materials* **35**, 2211598 (2023).

164. Guo, F. *et al.* Multifunctional Optoelectronic Synapse Based on Ferroelectric Van der Waals Heterostructure for Emulating the Entire Human Visual System. *Adv Funct Materials* **32**, 2108014 (2022).

165. Baek, S. *et al.* Ferroelectric Field-Effect-Transistor Integrated with Ferroelectrics Heterostructure. *Advanced Science* **9**, 2200566 (2022).

166. Soliman, M. *et al.* Photoferroelectric All-van-der-Waals Heterostructure for Multimode

Neuromorphic Ferroelectric Transistors. *ACS Appl. Mater. Interfaces* **15**, 15732–15744 (2023).